\documentclass[%
 aip,
 amsmath,amssymb,
 reprint,%
]{revtex4-1}

\usepackage{graphicx}
\usepackage{dcolumn}
\usepackage{bm}

\usepackage[utf8]{inputenc}
\usepackage[T1]{fontenc}
\usepackage{mathptmx}
\usepackage{ulem}
\usepackage[dvipsnames]{xcolor}

\begin{document}

\preprint{AIP/123-QED}

\title{Expansion of Ultracold Neutral Plasmas with Exponentially Decaying Density Distributions}

\author{M. K. Warrens}
 \affiliation{Rice University, 6100 Main St. Houston, TX, 77005}
\author{G. M. Gorman}
\affiliation{Rice University, 6100 Main St. Houston, TX, 77005}

\author{S. J. Bradshaw}
\affiliation{Rice University, 6100 Main St. Houston, TX, 77005}

\author{T. C. Killian}
\affiliation{Rice University, 6100 Main St. Houston, TX, 77005}

\date{\today}

\begin{abstract}
We present a study of the expansion of an  ultracold neutral plasma (UCNP) with an initial  density distribution that decays exponentially in space, created by photoionizing atoms shortly after their release from a quadrupole (or biconic cusp) magnetic trap. A characteristic ion acoustic timescale is evident in the evolution of the plasma size and velocity, indicating the dynamics are reasonably well described by a model of hydrodynamic expansion of a quasi-neutral plasma. However, for low plasma density and high initial electron temperature, excess ion kinetic energy in the vicinity of the central density peak suggests significant local non-neutrality at early times. Observations are compared to the well-understood self-similar expansion of an UCNP with an initial Gaussian density distribution, and a similar scaling law describes the evolution of plasma size for both cases.

\end{abstract}

\maketitle

\section{\label{sec:intro}Introduction}
Expansion of plasma into surrounding vacuum is a common phenomenon in many situations, such as astrophysical  \cite{kde87,cga04} and solar plasmas, \cite{pdf13,bra13,wbc08}  and plasmas created by intense laser-matter interaction. \cite{ckz00, skh00, mgf00, bwp01, fsk05, rfb05, mkn05, mor05physplasmas, gmo06, shh07} Depending upon plasma length scales and particle velocities, complex phenomena can arise reflecting hydrodynamic and kinetic effects, collective modes, and instabilities. 
Here, we investigate the expansion of an unmagnetized ultracold neutral plasma (UCNP) with an initial density distribution that is sharply peaked and decays exponentially with increasing distance from the plasma center. 

Expansion of an UCNP with a spherical Gaussian density distribution is well-studied \cite{kkb00,rha02,rha03,lgs07} and known to be self-similar and predominantly hydrodynamic. The sharply peaked density profile of the exponentially decaying distribution raises questions about the validity of a hydrodynamic description in this case.  Understanding the dynamics for this new plasma shape is important because ongoing experiments studying magnetized UCNPs in biconic cusp magnetic fields \cite{gwb20}   typically start with a plasma with an exponentially decaying density distribution. In order to identify the influence of magnetic forces, it is beneficial to understand the unmagnetized case.


\section{\label{sec:UCNPs}UCNPs}

UCNPs \cite{kkb99,kpp07,lro17} are formed by photo-exciting laser cooled atoms \cite{mvs99} or molecules in a supersonic expansion \cite{mrk08} near the ionization threshold, yielding ion temperatures  below $1$\,K, electron temperatures ranging from $1-1000$\,K, and densities ranging from $10^6-10^{12}$\,cm$^{-3}$. Such systems have shed light on equilibration \cite{scg04,ccb05,mur01,kon02,mck02,gms03,mur06PRL,mur07PoP,ppr04PRA,ppr05PRL,csl04,ccb05,lcg06,bcm12,lbm13} and transport \cite{slm16} in strongly coupled plasmas, \cite{ich82} which has connections to  laser-produced plasmas \cite{mur06PRL,lbh15} extending into the  warm-dense-matter and high-energy-density regimes. \cite{mur04,bbl19} With excellent control over the initial conditions and precise diagnostics, they have been used to benchmark theories and numerical techniques for quantum, \cite{glw20} molecular-dynamic, \cite{bcm12,slm16} kinetic, \cite{ppr04PRA} and hydrodynamic descriptions, \cite{mcs13,mcb15} and to observe and characterize collective modes and instabilities. \cite{kkb00,fzr06,zfr08,cmk10} UCNPs are also being developed for applications in bright ion and electron beams. \cite{cgt05,rkt09,mss11,msk16,fkv18,frn19} UCNPs created from molecules display additional processes such as dissociative recombination \cite{sgr12} and arrested relaxation.\cite{hss17}

For the experiments discussed here, UCNPs are formed by direct photoionization of laser-cooled Sr atoms. \cite{scg04,lgk19} 
After photoionization, an UCNP undergoes several stages of equilibration and expansion. The first stage is electron equilibration on the timescale of an electron plasma oscillation. Direct ionization yields a well-defined electron thermal energy given by the energy excess of the ionizing laser above the ionization threshold.\cite{kkb99} For low electron energy or high density, heating processes such as disorder-induced heating  and three-body recombination \cite{mke69, klk01} may  increase the electron temperature. \cite{kon02,rha02,gls07} This regime is avoided in the experiments described here. During electron equilibration, some of the electrons escape, creating a potential energy well that traps the remaining electrons. \cite{kkb99} 
The second stage of the UCNP evolution is ion equilibration, which occurs on the timescale of an ion plasma oscillation, which is $\sim 1\,\mu$s for typical UCNP densities. Disorder-induced heating causes the ion temperature to increase as the initially uncorrelated ions develop spatial correlations and convert excess potential energy to kinetic energy. \cite{mur01,csl04} The final stage of the UCNP evolution is expansion into surrounding vacuum.\cite{pmo94,lgs07}

       
        

   

The plasma expansion depends critically  upon the plasma's initial  density distribution, which follows the density profile of the neutral atoms just before photoionization. For UCNPs created from laser-cooled atoms, atoms are typically photoionized after release from a magneto-optical trap (MOT), a trap formed by near-resonant light scattering in the presence of spatially varying Zeeman shifts of the optical transition. \cite{mvs99, rpc87} This results in a  Gaussian density distribution of

\begin{equation}
\label{eq:GaussianDensityDensityDistribution}
    n(\vec{r}) = n_0\,\mathrm{exp}\Bigg [ -\frac{x^2}{2\sigma_x^2}-\frac{y^2}{2\sigma_y^2}-\frac{z^2}{2\sigma_z^2}\Bigg ].
\end{equation}

An alternative scheme is to trap atoms in a purely magnetic trap before photoionization. Such a trap is formed by the conservative potential $U=-\vec{\mu} \cdot \vec{B}$ arising from the interaction of an atom's magnetic moment $\vec{\mu}$ with the magnetic field $\vec{B}$. \cite{mvs99} Magnetic trapping is readily available for Sr atoms, \cite{nsl03} and as described in Sec.\ \ref{sec:ExpMethods}, the particular field geometry used for Sr UCNP experiments results in a cylindrically symmetric, exponentially decaying density for the atomic cloud and plasma,
\begin{equation}
n(\vec{r}) = n_0\mathrm{exp}\Bigg [ -\frac{\sqrt{x^2+ ({y^2} + {z^2})/{4}\eta^2}}{\alpha} \Bigg ].
\label{eq:cuspyDensityDensityDistribution}
\end{equation}
$\eta=1$ for an atomic sample in thermal equilibrium, but we allow for this parameter to vary as an additional degree of freedom when analyzing data.

 For the rest of the paper, we will refer to plasmas with Gaussian (Eq.\ \ref{eq:GaussianDensityDensityDistribution}) and exponentially decaying (Eq.\ \ref{eq:cuspyDensityDensityDistribution}) initial density distributions as Guassian plasmas and exponential plasmas respectively.

 \section{\label{sec:ExpMethods}Experimental Methods}
 
 \subsection{Details of Sr Ultracold Neutral Plasma Creation}
\begin{figure}[!ht]
    \centering
   \includegraphics[width = 0.40\textwidth]{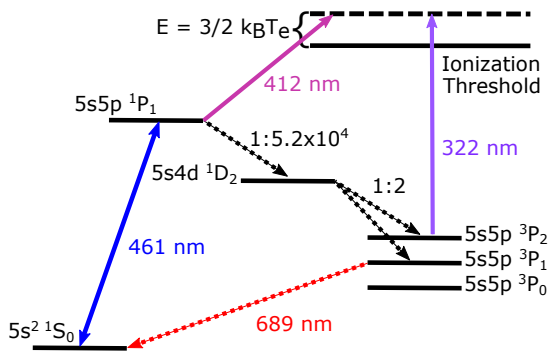}
   \caption{\small{Levels and  transitions for  laser cooling of strontium atoms and subsequent photoionization to form an UCNP.}}
   \label{fig:atomlevels}
\end{figure}

To form Gaussian UCNPs, Sr atoms are first laser-cooled and trapped in a MOT  using the $^1$S$_0-^1$P$_1$ transition at 461 nm (Fig.\ \ref{fig:atomlevels}).  The MOT is then extinguished and  atoms in the $5s^2$\,$^1$S$_0$ ground state are photoionized  using overlapping 10 ns pulses of 461 nm and $\sim 412$\, nm light. \cite{scg04} 
  
To form exponential UCNPs,  atoms are initially laser-cooled and trapped in the MOT. During laser cooling, atoms naturally populate the $5s5p$\,$^3$P$_2$ state due to a small branching ratio for decay from the upper state ($5s5p$\,$^1$P$_1$) through the $5s4d$\,$^1$D$_2$ state into the $5s5p$\,$^3$P$_2$ state, where low field seeking atoms are trapped by the quadrupole (or biconic cusp) magnetic field  of the MOT because of the large magnetic moment of $5s5p$\,$^3$P$_2$ atoms. \cite{nsl03}  The density distribution of the magnetically trapped neutral atoms is well approximated by Eq.\ \ref{eq:cuspyDensityDensityDistribution}. After loading the magnetic trap for a variable time $\sim 1$\,s, the magnetic field and laser-cooling light are turned off. After a delay of 280 $\,\mu$s to allow eddy currents in the vacuum chamber to decay,  $^3$P$_2$ atoms are photoionized with a 10 ns, 10 mJ pulse of $\sim 322$\,nm light.  This delay time is short compared to the timescale for appreciable motion of the atoms, and the ionizing beam is approximately uniform over the spatial extent of the atom cloud. By varying the loading time, the initial plasma density is varied from $1-40\times 10^{8}$\,cm$^{-3}$.

  
The characteristic size of the exponentially decaying plasma is $\alpha =3k_BT/8\mu_B B' \approx 1$ mm, where $T\approx 2$ mK is the temperature of the precursor neutral atoms, and $B'=150$ G/cm is the gradient of the magnetic field along the symmetry axis  ($\hat{x}$). $\mu_B$ and $k_B$ are the Bohr magneton and Boltzmann constant respectively.  The atom cloud is small compared to the radius of the coils creating the magnetic field, so a linear approximation of the field profile is sufficient. 

 
 

\subsection{Imaging an Ultracold Neutral Plasma}
We image  ions in an UCNP  using laser-induced fluorescence (LIF). \cite{cgk08} A 1 mm thick sheet of  422 nm light passes  through the center of the plasma and along the symmetry axis of the coils  generating the MOT magnetic field. This light drives the $5s\,\,^2$S$_{1/2}-5s\,\,^2$P$_{1/2}$ Sr$^+$ transition, and emitted fluorescence is imaged onto an intensified CCD camera along a direction ($\hat{z}$) perpendicular to the laser sheet. Varying the detuning of the imaging beam yields  a spatially resolved LIF spectrum.  Local ion temperature and velocity are extracted from a fit of the spectrum to  a Voigt profile. We image the plasma for 500 ns, which is short compared to the expansion time of the UCNP. Times reported are from the middle of the imaging window.   Density maps of the central plane of the plasma are created by integrating the local spectra over frequency. The proportionality between LIF intensity and density is calibrated using disorder-induced heating curves as an absolute measure of density. \cite{lsm16} 
The LIF optical configuration does not provide information on the $z$-dependence of the density distribution. In the analysis, we assume cylindrical symmetry about the $x$-axis, reflecting the magnetic field symmetry.

\begin{figure}[!ht]
    \centering
    \includegraphics[width = 0.45\textwidth]{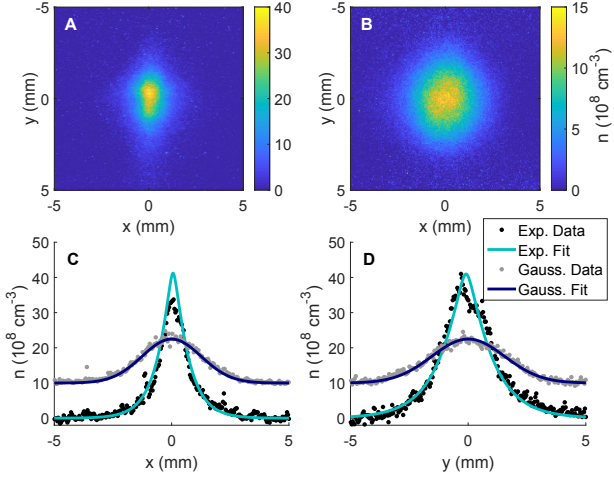}
    \caption{ (A) False-color image of the density profile at $t = 350$\,ns after photoionization for an exponential plasma with $T_e(0)=160$\, K. (B) Density profile at $t= 330$\,ns for a Gaussian plasma with $T_e(0)=40$\, K. (C) X-transect through the center of the plasmas shown in (A) and (B). (D) Y-transect through the center of the plasmas shown in (A) and (B). Fit lines in (C) and (D) are integrals over LIF-laser-sheet thickness of Eq. \ref{eq:GaussianDensityDensityDistribution} for the Gaussian plasma and  of Eq. \ref{eq:cuspyDensityDensityDistribution}  for the exponential plasma. 
    }
    \label{fig:ProofOfCuspyDensity}
\end{figure}

Figure \ref{fig:ProofOfCuspyDensity} shows false-color images and transects of the density for an exponential plasma with initial peak density $3.2\times 10^{9}$ cm$^{-3}$ and $T_e(0)$ = 160 K  and a Gaussian plasma with initial peak density $1.2 \times 10^{9}$ cm$^{-3}$ and $T_e(0)$ = 40 K shortly after photoionization. 
The fit function for the exponential plasma (Fig.\ \ref{fig:ProofOfCuspyDensity}) is obtained by integrating Eq.\ \ref{eq:cuspyDensityDensityDistribution} over the width of the imaging sheet, and $\alpha$, $\eta$ and $n_0$ are varied to match the data. We find $\eta=0.83$, close to the expected value.

Deviations between the data and fit most likely result from non-uniformities in the photoionization beam, lack of thermal equilibrium of atoms in the magnetic trap, and plasma evolution during the short delay between photoionization and LIF imaging. In spite of these imperfections, from the comparison of the Gaussian and exponential data in Fig. \ref{fig:ProofOfCuspyDensity}, it is clear that that the exponential plasma is much more sharply peaked than the Gaussian plasma.

\section{\label{sec:hydrodynamic} Plasma Expansion}

Hydrodynamic models accurately describe the evolution of plasma density in collisional systems, for which the length scales of plasma inhomogeneities are much larger than the particle mean free path and the particles are close to local thermal equilibrium. \cite{mcs13,mcb15} Assuming quasineutrality \cite{kpp07} ($n_e \approx n_i \equiv n$ for electron and  ion  densities  $n_e$ and $n_i$) and the absence of any magnetic forces, this implies  the following equation for the rate of change of the local hydrodynamic velocity  $\vec{u}$ in the limit of $T_i \ll T_e$
 
\begin{equation}
  m_i\dot{\vec{u}} \approx -k_BT_e\frac{\vec{\nabla}n}{n},
    \label{eq:hydrodynamicEOM}
\end{equation}
where $m_i$ is the ion mass, $k_B$ is the Boltzmann constant, and $T_e$ and $T_i$ are the electron and ion temperatures respectively. Eq. \ref{eq:hydrodynamicEOM} describes a hydrodynamic expansion driven by the electron pressure gradient.



\subsection{Self-similar Expansion of an UCNP with a Spherical Gaussian Density Distribution}

For Gaussian UNCPs, a hydrodynamic description in terms of coupled equations for electrons and ions captures the dominant behavior of UCNP expansion. \cite{kkb00,lgs07,kpp07} 
Various authors have shown that a hydrodynamic treatment neglecting ion-electron thermalization \cite{msl15} and inelastic processes \cite{mke69,gls07,pvs08} predicts that a Gaussian UCNP undergoes self-similar expansion, described by Eq.\ \ref{eq:GaussianDensityDensityDistribution} with $\sigma\rightarrow \sigma(t)$.  \cite{rha02,ppr04PRA,kpp07} A comprehensive treatment, \cite{ppr04PRA,kpp07} deriving the evolution of hydrodynamic moments from kinetic equations, yields the following equations describing the evolution of plasma density and constituent temperatures,

\begin{eqnarray}
\label{expansiona}
\sigma^2(t)&=&\sigma^2(0)\left(1+t^2/\tau_{\rm exp}^2\right)\;, \\
\label{expansionb}
\gamma(t)&=&\frac{t/\tau_{\rm exp}^2}{1+t^2/\tau_{\rm exp}^2}\;, \\
\label{expansionc}
T_{\rm i}(t)&=&\frac{T_{\rm i}(0)}{1+t^2/\tau_{\rm exp}^2}\;, \\
\label{expansiond}
T_{\rm e}(t)&=&\frac{T_{\rm e}(0)}{1+t^2/\tau_{\rm exp}^2}\;,
\end{eqnarray}
The characteristic plasma expansion time is given by
\begin{equation}
  \label{expansiontime}
    \tau_{\rm exp}=\sqrt{\frac{m_{\rm_i}\sigma(0)^2}{k_{\rm B} [T_{\rm e}(0)+T_{\rm i}(0)]}}\approx \sqrt{\frac{m_{\rm_i}\sigma(0)^2}{k_{\rm B} T_{\rm e}(0)}}\,
\end{equation}
and the
hydrodynamic expansion velocity follows 
\begin{equation}
    \label{expansionvelocity}
    \mathbf{u}(\mathbf{r},t)=\gamma(t) \mathbf{r}.
\end{equation}
Equations \ref{eq:hydrodynamicEOM} and \ref{expansionvelocity} both yield the acceleration  $\dot{\vec{u}} \approx \frac{k_BT_e(t)}{m_i\sigma(t)^2}\vec{r}$. The linear dependence of the velocity and acceleration upon displacement vector $\vec{r}$ ensures the self-similarity of the expansion. Experiments confirm that  the analytic solution describes the expansion of a Gaussian UCNP extremely well. \cite{lgs07}

\subsection{Expansion of an UCNP with an Exponentially Decaying Density Distribution}

\begin{figure}[!ht]
    \centering
    \includegraphics[width = 0.45\textwidth]{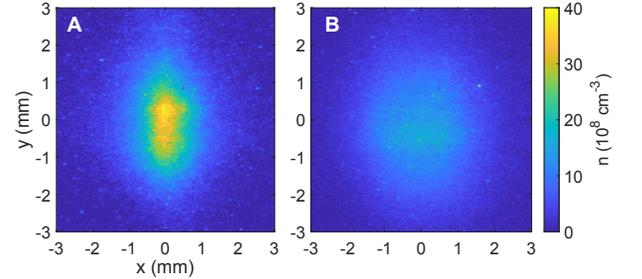}
    \caption{False color images showing expansion of an exponential plasma with initial peak density $4.4\times 10^{9}$\,cm$^{-3}$ and $T_e(0)=40$\, K. (A) $t=350$\,ns  after plasma creation. (B) $t=11$\,$\mu$s after plasma creation. }
    \label{fig:aspectRatioInversion40K}
\end{figure}

\begin{figure*}[t]
    \centering
    \includegraphics[width = \textwidth]{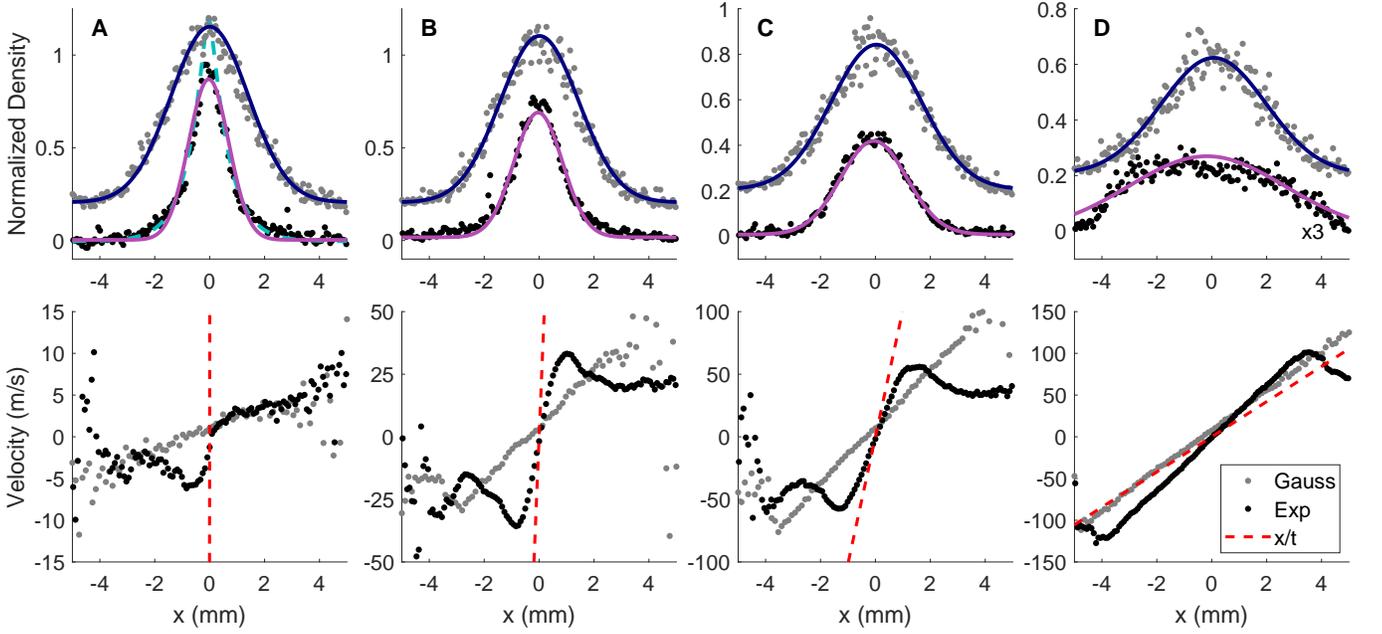}
    \caption{Transects along the $x$-axis of the  plasma density distribution (top) and velocity (bottom), for Gaussian and exponential plasmas for initial peak densities of $1.0 \times 10^{9}$\,cm$^{-3}$ and $3.5 \times 10^{9}$\,cm$^{-3}$ respectively and  electron temperature  $T_e(0)=40$\, K. Density transects for the Gaussian plasmas are offset for clarity.   The time after photoionization is (A) 0.34 $\pm$ 0.01 $\mu s$, (B) 5.7 $\pm$ 0.1 $\mu s$, (C) 14 $\pm$ 4 $\mu s$,  and (D) 27 $\pm$ 3 $\mu s$. All density transects are fit to integrals over LIF-laser-sheet thickness of Eq. \ref{eq:GaussianDensityDensityDistribution} (solid lines). The first time point (A) for the exponential plasma is also fit to the integral over LIF-laser-sheet thickness of  Eq.\ \ref{eq:cuspyDensityDensityDistribution} (dashed line).}
    \label{fig:XTransects}
\end{figure*}

For an exponential plasma (Eq.\ \ref{eq:cuspyDensityDensityDistribution}), the length scale for density variation is $|{n}/{\vec{\nabla}n}|\sim \alpha \sim 1$\,mm. The ion mean-free path is typically several orders of magnitude smaller than this, and the electron mean-free path is typically similar to or smaller than $\alpha$ for most of the plasma over the range of accessible UCNP conditions. In addition, electrons rapidly thermalize with small Debye screening length in the self-consistent potential of the ions and electrons. This suggests that hydrodynamic models should describe the expansion well, as observed in previous experiments with Gaussian plasmas, \cite{lgs07,mcs13} except perhaps very near the center of the plasma where the density gradient is not well defined.

A first indication of the hydrodynamic nature of expansion of an exponential UCNP is given by the false-color images of the density (Fig.\ \ref{fig:aspectRatioInversion40K}). The rapid expansion along the initially narrow axis of the plasma is a characteristic signature of hydrodynamic behavior in which the acceleration is proportional to the local density gradient (Eq.\ \ref{eq:hydrodynamicEOM}).

To go beyond this qualitative conclusion, we look quantitatively at the evolution of the plasma size. Figure \ref{fig:XTransects}(top) shows transects  of the evolving density distributions for Gaussian   and exponential plasmas with initial electron temperature of $T_e(0)=40$\,K. The transects pass through the plasma center along the initially smaller dimension  of the exponential plasma (along the \textit{x}-axis). Density is normalized by the initial peak density.  The Gaussian plasma remains Gaussian as predicted for the self-similar expansion (Eqs. \ref{expansiona}-\ref{expansiond}). \cite{lgs07}  The transects for the exponential plasma show that the plasma is initially exponential, but does not expand self-similarly. The density first decreases at the center, rounding out the sharp peak and causing the density distribution to approach a Gaussian. However, it does not retain the Gaussian shape, and the gradient steepens along the advancing edges of the plasma, suggestive of wave breaking. \cite{mor05physplasmas,rha03}

\begin{figure}[t]
\includegraphics[width=.45\textwidth]{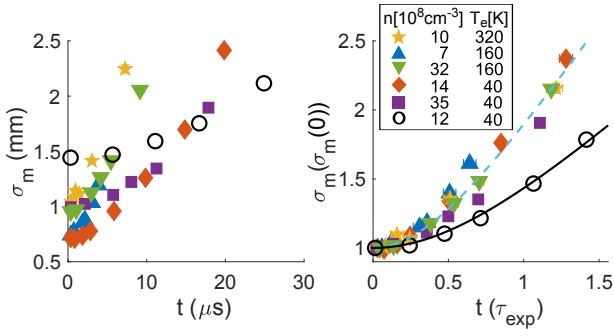}
\caption{\label{fig:SigVsTime} Evolution of the RMS mean plasma size ($\sigma_m$) with time. Open and closed symbols correspond to Gaussian and exponential plasmas  respectively. Initial peak density and electron temperature is indicated in the legend.  Size and time are unscaled (left) and scaled (right) by initial size and expansion time (Eq. \ref{expansiontime}) respectively.  Fit lines correspond to  Eq. \ref{eq:modifiedSigma} with $\beta=1$ (solid) and $\beta=0.63$ (dashed).
}
\end{figure}

To quantitatively describe the plasma expansion, we extract characteristic sizes with numeric fits to the data (Fig. \ref{fig:XTransects}(top)). The  exponential (Eq.\ \ref{eq:cuspyDensityDensityDistribution}), integrated over the LIF laser width, accurately describes the exponential plasma initially, but fails as the plasma evolves. 
 To characterize the size evolution, we fit the density transects of both Gaussian and exponential plasmas to integrals over LIF-laser-sheet thickness of the Gaussian expression (Eq. \ref{eq:GaussianDensityDensityDistribution}).
The RMS radii $\sigma_x$ and $\sigma_y$ are extracted from the fits. Although the exponential plasma shape is not well described by the Gaussian for some time points, this function captures the size of the plasma throughout the evolution.  We assume cylindrical symmetry about the x-axis for all data, $\sigma_y = \sigma_z$.  Figure \ref{fig:SigVsTime} shows the evolution of the geometric mean plasma size $\sigma_{m} = (\sigma_x\sigma_y\sigma_z)^{1/3} = (\sigma_x\sigma_y^2)^{1/3}$ for a Gaussian plasma with $T_e(0) = 40$ K and several exponential plasmas of various $T_e(0)$.






For the exponential plasmas, it is clear from the plot with unscaled axes (Fig.\ \ref{fig:SigVsTime}A) that smaller initial mean size and higher $T_e(0)$ leads to faster expansion. To highlight this, Fig.\ \ref{fig:SigVsTime}B  shows the data with size scaled by the initial value and time scaled by $\tau_{exp}$ (Eq.\ \ref{expansiontime}), where we have set $\sigma(0)$ equal to the initial fit $\sigma_m$.
In spite of the fact that the exponential plasmas lack spherical symmetry and Gaussian curvature, and thus do not satisfy the initial conditions of the self-similar expansion, this scaling causes the data to collapse to close to a universal curve over this parameter range. This is another strong indication that  the expansion is predominantly hydrodynamic in nature. It also  suggests a very useful phenomenological description of the plasma expansion.

Figure \ref{fig:SigVsTime}B shows that the expansion of the Gaussian plasma is well-described by the universal curve from Eq. \ref{expansiona}. To fit the exponential data we use Eq. \ref{expansiona}, but with a modified effective expansion time $\beta \tau_{exp}$, 
\begin{equation}
    \sigma_m(t)/\sigma_m(0) = \sqrt{1 + t^2/(\beta \tau_{exp})^2}
    \label{eq:modifiedSigma}
\end{equation}
We find $\beta \approx 0.63$ for the range of conditions covered here, and $\tau_{exp}$ is calculated from Eq.\ \ref{expansiontime} using the initial mean size $\sigma_m(0)$. This provides a valuable and intuitive phenomenological description of the plasma expansion. The scaling for exponential plasmas is not perfect in Fig.\ \ref{fig:SigVsTime}B. Plasmas with higher $T_e(0)$ and lower density expand slightly faster in scaled units. This trend is more evident if one fits the data sets for each initial condition individually independent values of $\beta$. This suggests a small influence of kinetic effects,\cite{mcb15} such as free-streaming, which enhances density at the expanding fronts and broadens the width of the fitted curve at each scaled time.

\section{\label{sec:velocity} Velocity Evolution}

Figure \ref{fig:XTransects}(bottom) shows the $x$-component of the hydrodynamic velocity $\vec{u}$ along an $x$-transect passing through the center of the plasma for the same data as Fig.\ \ref{fig:XTransects}(top). At all times,  the Gaussian plasma displays the characteristic linear increase of the velocity with distance from the plasma center (Eq.\ \ref{expansionvelocity}) that is a hallmark of the self-similar expansion. Linearity extends essentially until the density is too low to obtain a reliable measure of velocity, beyond $r\sim 2 \sigma_m(0)$, where the density is $\lesssim 10$\% of its peak value. The  velocity curve  approaches $\vec{u}=r/t$ for $t \gg \beta\tau_{exp}$.


The velocity profile for  the exponential plasma  deviates dramatically from the prediction of the self-similar solution  (Eq.\ \ref{expansionvelocity}) as shown by Fig. \ref{fig:XTransects}(bottom). For an exponential density profile (Eq. \ref{eq:cuspyDensityDensityDistribution}), the acceleration  along the x-axis due to the electron pressure gradient is $\dot{u}_x = sgn(x){k_BT_e}/{m_i\alpha}$. Even with   smoothing of the central density peak in the initial distribution due to experimental imperfections (Fig.\ \ref{fig:ProofOfCuspyDensity}(C-D)), this form of the acceleration suggests that a steep velocity gradient should form very quickly near the plasma center, as evident in Fig. \ref{fig:XTransects}(bottom).
Further from the center, there are extrema in the velocity after just a few microseconds. As the plasma evolves, the extrema move out, and they cause the steepening of the density gradient observed at later times in Fig.\ \ref{fig:XTransects}(top).

\begin{figure}[!ht]
{\includegraphics[width=0.48\textwidth]{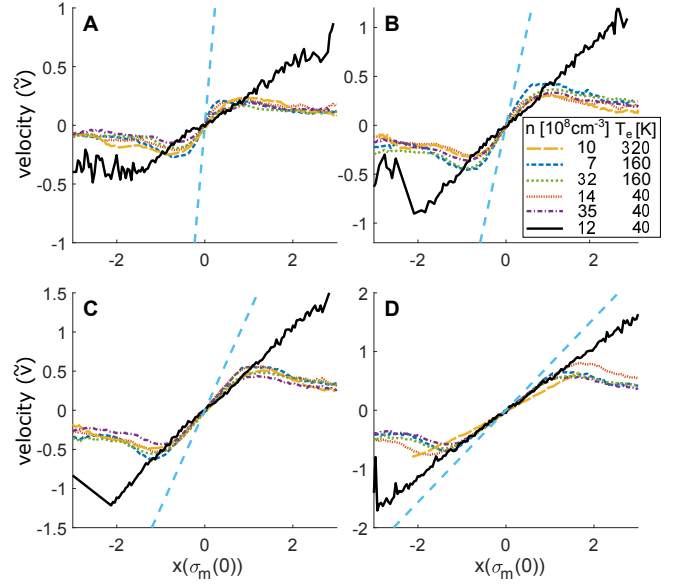}}
\caption{\label{fig:scaledVelocitysTransects} Transects passing through the plasma origin along the $x$-axis of the $x$-component of the hydrodynamic velocity at $t/\beta\tau_{exp} = $  0.23 $\pm$ 0.02 (A), 0.49 $\pm$ 0.09 (B),  0.80 $\pm$ 0.05 (C),  1.28 $\pm$ 0.35 (D).   Velocity is given in units of $\tilde{v}= \sigma_m(0)/\beta \tau_{exp}=\sqrt{k_BT_e(0)/m_i}/\beta$, and position is scaled by $\sigma_m(0)$. Initial  peak density and $T_e(0)$ are indicated in the legend. The long dashed line corresponds to the terminal velocity profile $\vec{u}=\vec{r}/t$. The solid line is a Gaussian plasma. All other data corresponds to  exponential plasmas.}
\end{figure}

 
The hydrodynamic nature of the expansion for the exponential plasma is  evident in the universality of the velocity profiles in appropriate scaled units, in analogy to the behavior of the self-similar hydrodynamic expansion (Eqs.\ \ref{expansiona}-\ref{expansionvelocity}). Figure \ref{fig:scaledVelocitysTransects} shows the $x$-component of expansion velocity  along the x-axis for various initial conditions at the same times in units of $\beta\tau_{exp}$. Velocity and position are scaled in units of  $\sigma_m(0)/\beta\tau_{exp}$  and $\sigma_m(0)$ respectively.  In scaled units, the velocity curves for exponential plasmas are universal, with deviations mostly due to spread in values of $t/\beta\tau_{exp}$ that are displayed in each subplot. The velocity extrema  approach $v\approx \sqrt{k_BT_e(0)/m_i}/\beta$. 
At early times and  near the plasma center, the density gradient, acceleration, and scaled velocity is much higher for the  exponential plasmas  than for Gaussian plasma.
The velocity in scaled units for the Gaussian plasma  uniformly and monotonically approaches the ballistic limit, as expected from Eq.\ \ref{expansionvelocity}, and eventually matches the exponential plasma in the center.

\section{\label{sec:temp} Temperature Evolution}

\begin{figure}[!ht]
    \includegraphics[width=.45\textwidth]{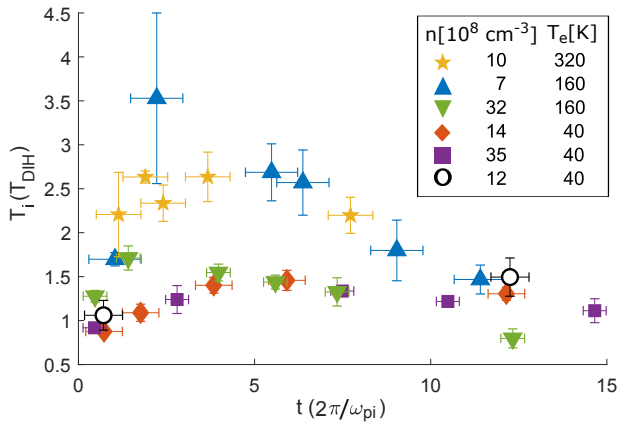}
    \caption{\label{fig:TempVsTime} Evolution of the ion temperature in the plasma center. Temperatures are scaled by the disorder-induced-heating temperature $T_{DIH}$ and time is scaled by the ion oscillation period $2\pi/\omega_{pi}$. The Gaussian plasma equilibrates to $T_{DIH}$ within $2\pi/\omega_{pi}$, as expected from disorder-induced heating. This is also seen with the exponential plasmas for $T_e =40$ K. However, for the exponential plasmas with $T_e = 160, \, 320$ K, significant heating is seen, especially for the $T_e = 320$ K and the low density $T_e =160$ K data, indicative of kinetic effects from a Coulomb explosion.}
\end{figure}

While a hydrodynamic description assuming quasi-neutrality (Eq.\ \ref{eq:hydrodynamicEOM}) appears to capture the behavior of the exponential plasma very well,  the assumption of quasi-neutrality is suspect near the plasma center. The Debye length for electrons, $\lambda_D=\sqrt{\varepsilon_0 k_B T_e/n_e e^2}$ for $n_e=10^{9}$\,cm$^{-3}$ and $T_e=320$\,K  is $40\,\mu$m, which is on the order of the length scale of the central peak, so significant non-neutrality is possible near the origin for high-temperature and low-density exponential plasmas. Figure \ref{fig:TempVsTime} shows the evolution of the ion temperature in the center of the plasma, which displays effects likely arising from non-neutrality.

The evolution of ion temperature for exponential plasmas for $T_e(0)=40$\,K and for the Gaussian plasma follows the expected behavior for ion equilibration  after UCNP creation, which is well studied.  \cite{mur01,scg04,lbm13,msl15,lsm16} Ions have very little initial kinetic energy because of the controlled UCNP creation process. However, there is a relatively large amount of potential energy due to electrical interactions between ions in the random spatial distribution  inherited from the precursor atoms. In a time equal to the inverse of the ion plasma oscillation frequency, $1/\omega_{pi}=(e^2n/\epsilon_0 m_i)^{-1/2} \sim 1\,\mu$s, ions develop spatial correlations, and potential energy is changed into kinetic as the system equilibrates. The equilibration temperature, defined as $T_{DIH}$, depends on initial density and electron temperature. Assuming overall plasma neutrality, $T_{DIH}$ can be  directly found with molecular dynamics \cite{lsm16} or expressed in terms of analytic approximations of equilibrium correlation energy determined with molecular dynamics simulations. \cite{lbm13,hfd97} $T_{DIH}\approx e^2/(8 \pi \epsilon_0 a k_B)\sim 1$\,K  where $a=\left[3/(4\pi n)\right]^{1/3}$ is the Wigner-Seitz radius. Figure \ref{fig:TempVsTime} shows that for low $T_e(0)$, ions equilibrate with $T/T_{DIH}\sim 1$. Subsequent moderate heating on a longer timescale reflects dissipation of ion acoustic waves and \cite{cmk10,mcb15} and electron-ion energy exchange. \cite{msl15}

For $T_e(0)=160$\,K and $T_e(0)=320$\,K, and more dramatically for lower ion density, however, ions reach much higher scaled temperatures on a rapid timescale. We hypothesize that this extra heating arises from plasma non-neutrality, which introduces additional Coulomb energy in the central region and subsequent Coulomb explosion.

\section{\label{sec:conclusion} Conclusion}
We characterized the expansion of an UCNP with an initial exponentially decaying density distribution and compared it to the well-studied self-similar expansion of an UCNP with a Gaussian density distribution. The expansion of an exponential UCNP  displays the hallmarks of hydrodynamic phenomenon and is driven by the electron pressure gradient. The characteristic length ($\sigma_0$) and time  scale ($\tau_{exp}$) that parameterize the  expansion of Gaussian plasmas  also provide useful scaling factors for evolution of the size and velocity distribution for exponential plasmas. 

The ion temperatures display excess heating near the density peak in the exponential case for low density and high electron temperature, suggesting significant non-neutrality that might introduce dynamics similar to a Coulomb explosion in this region. Further study of this phenomena is needed.

Modeling the expansion of an exponential UNCP is an excellent target  for various numerical methods, such as two-fluid models, models that consider kinetic effects, and MD simulations. It would be informative to compare the observations described here with two-component MD codes with open boundary conditions \cite{vbk20}
and recently developed tools for describing heterogeneous, non-equilibrium plasmas subject to effects of strong coupling. \cite{dvm20}  This would provide validation of numerical methods and additional insight into some aspects of the expansion of the exponential plasma, such as electron temperature evolution and the influence of Coulomb-explosion effects on ion temperature near the plasma center. Further study is also needed to explore the possible formation of shock waves associated with the extrema in the velocity transects and steepening of the density gradient.



\begin{acknowledgments}
This material is based upon work supported by the Air Force Office of Scientific Research through Grant No. FA9550-17-1-0391 and the National Science Foundation Graduate Research Fellowship Program under Grant No. 1842494. Any opinions, findings, and conclusions or recommendations expressed in this material are those of the authors and do not necessarily reflect the views of the National Science Foundation.
\end{acknowledgments}

\section*{Data Availability}
The data that support the finding of this study are available from the corresponding author upon reasonable request.

\bibliography{bibliography}

\begin{thebibliography}{72}%
\makeatletter
\providecommand \@ifxundefined [1]{%
 \@ifx{#1\undefined}
}%
\providecommand \@ifnum [1]{%
 \ifnum #1\expandafter \@firstoftwo
 \else \expandafter \@secondoftwo
 \fi
}%
\providecommand \@ifx [1]{%
 \ifx #1\expandafter \@firstoftwo
 \else \expandafter \@secondoftwo
 \fi
}%
\providecommand \natexlab [1]{#1}%
\providecommand \enquote  [1]{``#1''}%
\providecommand \bibnamefont  [1]{#1}%
\providecommand \bibfnamefont [1]{#1}%
\providecommand \citenamefont [1]{#1}%
\providecommand \href@noop [0]{\@secondoftwo}%
\providecommand \href [0]{\begingroup \@sanitize@url \@href}%
\providecommand \@href[1]{\@@startlink{#1}\@@href}%
\providecommand \@@href[1]{\endgroup#1\@@endlink}%
\providecommand \@sanitize@url [0]{\catcode `\\12\catcode `\$12\catcode
  `\&12\catcode `\#12\catcode `\^12\catcode `\_12\catcode `\%12\relax}%
\providecommand \@@startlink[1]{}%
\providecommand \@@endlink[0]{}%
\providecommand \url  [0]{\begingroup\@sanitize@url \@url }%
\providecommand \@url [1]{\endgroup\@href {#1}{\urlprefix }}%
\providecommand \urlprefix  [0]{URL }%
\providecommand \Eprint [0]{\href }%
\providecommand \doibase [0]{http://dx.doi.org/}%
\providecommand \selectlanguage [0]{\@gobble}%
\providecommand \bibinfo  [0]{\@secondoftwo}%
\providecommand \bibfield  [0]{\@secondoftwo}%
\providecommand \translation [1]{[#1]}%
\providecommand \BibitemOpen [0]{}%
\providecommand \bibitemStop [0]{}%
\providecommand \bibitemNoStop [0]{.\EOS\space}%
\providecommand \EOS [0]{\spacefactor3000\relax}%
\providecommand \BibitemShut  [1]{\csname bibitem#1\endcsname}%
\let\auto@bib@innerbib\@empty
\bibitem [{\citenamefont {Karpen}\ and\ \citenamefont {DeVore}(1987)}]{kde87}%
  \BibitemOpen
  \bibfield  {author} {\bibinfo {author} {\bibfnamefont {J.~T.}\ \bibnamefont
  {Karpen}}\ and\ \bibinfo {author} {\bibfnamefont {C.~R.}\ \bibnamefont
  {DeVore}},\ }\href@noop {} {\bibfield  {journal} {\bibinfo  {journal}
  {{ApJ}}\ }\textbf {\bibinfo {volume} {320}},\ \bibinfo {pages} {904}
  (\bibinfo {year} {1987})}\BibitemShut {NoStop}%
\bibitem [{\citenamefont {Courtois}\ \emph {et~al.}(2005)\citenamefont
  {Courtois}, \citenamefont {Grundy}, \citenamefont {Ash}, \citenamefont
  {Chambers}, \citenamefont {Woolsey}, \citenamefont {Dendy},\ and\
  \citenamefont {McClements}}]{cga04}%
  \BibitemOpen
  \bibfield  {author} {\bibinfo {author} {\bibfnamefont {C.}~\bibnamefont
  {Courtois}}, \bibinfo {author} {\bibfnamefont {R.~A.~D.}\ \bibnamefont
  {Grundy}}, \bibinfo {author} {\bibfnamefont {A.~D.}\ \bibnamefont {Ash}},
  \bibinfo {author} {\bibfnamefont {D.~M.}\ \bibnamefont {Chambers}}, \bibinfo
  {author} {\bibfnamefont {N.~C.}\ \bibnamefont {Woolsey}}, \bibinfo {author}
  {\bibfnamefont {R.~O.}\ \bibnamefont {Dendy}}, \ and\ \bibinfo {author}
  {\bibfnamefont {K.~G.}\ \bibnamefont {McClements}},\ }\href@noop {}
  {\bibfield  {journal} {\bibinfo  {journal} {Phys. Plasmas}\ }\textbf
  {\bibinfo {volume} {11}},\ \bibinfo {pages} {3386} (\bibinfo {year}
  {2005})}\BibitemShut {NoStop}%
\bibitem [{\citenamefont {Perrone}\ \emph {et~al.}(2013)\citenamefont
  {Perrone}, \citenamefont {Dendy}, \citenamefont {Furno}, \citenamefont
  {Sanchez}, \citenamefont {Zimbardo}, \citenamefont {Bovet}, \citenamefont
  {Fasoli}, \citenamefont {Gustafson}, \citenamefont {Perri}, \citenamefont
  {Ricci},\ and\ \citenamefont {Valentini}}]{pdf13}%
  \BibitemOpen
  \bibfield  {author} {\bibinfo {author} {\bibfnamefont {D.}~\bibnamefont
  {Perrone}}, \bibinfo {author} {\bibfnamefont {R.~O.}\ \bibnamefont {Dendy}},
  \bibinfo {author} {\bibfnamefont {I.}~\bibnamefont {Furno}}, \bibinfo
  {author} {\bibfnamefont {R.}~\bibnamefont {Sanchez}}, \bibinfo {author}
  {\bibfnamefont {G.}~\bibnamefont {Zimbardo}}, \bibinfo {author}
  {\bibfnamefont {A.}~\bibnamefont {Bovet}}, \bibinfo {author} {\bibfnamefont
  {A.}~\bibnamefont {Fasoli}}, \bibinfo {author} {\bibfnamefont
  {K.}~\bibnamefont {Gustafson}}, \bibinfo {author} {\bibfnamefont
  {S.}~\bibnamefont {Perri}}, \bibinfo {author} {\bibfnamefont
  {P.}~\bibnamefont {Ricci}}, \ and\ \bibinfo {author} {\bibfnamefont
  {F.}~\bibnamefont {Valentini}},\ }\href {\doibase 10.1007/s11214-013-9966-9}
  {\bibfield  {journal} {\bibinfo  {journal} {Space Sci. Rev.}\ }\textbf
  {\bibinfo {volume} {178}},\ \bibinfo {pages} {233} (\bibinfo {year}
  {2013})}\BibitemShut {NoStop}%
\bibitem [{\citenamefont {Bradshaw}\ and\ \citenamefont
  {Raymond}(2013)}]{bra13}%
  \BibitemOpen
  \bibfield  {author} {\bibinfo {author} {\bibfnamefont {S.~J.}\ \bibnamefont
  {Bradshaw}}\ and\ \bibinfo {author} {\bibfnamefont {J.}~\bibnamefont
  {Raymond}},\ }\href {\doibase 10.1007/s11214-013-9970-0} {\bibfield
  {journal} {\bibinfo  {journal} {Space Sci. Rev.}\ }\textbf {\bibinfo {volume}
  {178}},\ \bibinfo {pages} {271} (\bibinfo {year} {2013})}\BibitemShut
  {NoStop}%
\bibitem [{\citenamefont {West}, \citenamefont {Bradshaw},\ and\ \citenamefont
  {Cargill}(2008)}]{wbc08}%
  \BibitemOpen
  \bibfield  {author} {\bibinfo {author} {\bibfnamefont {M.~J.}\ \bibnamefont
  {West}}, \bibinfo {author} {\bibfnamefont {S.~J.}\ \bibnamefont {Bradshaw}},
  \ and\ \bibinfo {author} {\bibfnamefont {P.~J.}\ \bibnamefont {Cargill}},\
  }\href {\doibase 10.1007/s11207-008-9243-3} {\bibfield  {journal} {\bibinfo
  {journal} {Sol. Phys.}\ }\textbf {\bibinfo {volume} {252}},\ \bibinfo {pages}
  {89} (\bibinfo {year} {2008})}\BibitemShut {NoStop}%
\bibitem [{\citenamefont {Clark}\ \emph {et~al.}(2000)\citenamefont {Clark},
  \citenamefont {Krushelnick}, \citenamefont {Zepf}, \citenamefont {Beg},
  \citenamefont {Tatarakis}, \citenamefont {Machacek}, \citenamefont {Santala},
  \citenamefont {Watts}, \citenamefont {Norreys},\ and\ \citenamefont
  {Dangor}}]{ckz00}%
  \BibitemOpen
  \bibfield  {author} {\bibinfo {author} {\bibfnamefont {E.~L.}\ \bibnamefont
  {Clark}}, \bibinfo {author} {\bibfnamefont {K.}~\bibnamefont {Krushelnick}},
  \bibinfo {author} {\bibfnamefont {M.}~\bibnamefont {Zepf}}, \bibinfo {author}
  {\bibfnamefont {F.~N.}\ \bibnamefont {Beg}}, \bibinfo {author} {\bibfnamefont
  {M.}~\bibnamefont {Tatarakis}}, \bibinfo {author} {\bibfnamefont
  {A.}~\bibnamefont {Machacek}}, \bibinfo {author} {\bibfnamefont {M.~I.~K.}\
  \bibnamefont {Santala}}, \bibinfo {author} {\bibfnamefont {I.}~\bibnamefont
  {Watts}}, \bibinfo {author} {\bibfnamefont {P.~A.}\ \bibnamefont {Norreys}},
  \ and\ \bibinfo {author} {\bibfnamefont {A.~E.}\ \bibnamefont {Dangor}},\
  }\href@noop {} {\bibfield  {journal} {\bibinfo  {journal} {Phys. Rev. Lett.}\
  }\textbf {\bibinfo {volume} {85}},\ \bibinfo {pages} {1654} (\bibinfo {year}
  {2000})}\BibitemShut {NoStop}%
\bibitem [{\citenamefont {Snavely}\ \emph {et~al.}(2000)\citenamefont
  {Snavely}, \citenamefont {Key}, \citenamefont {Hatchett}, \citenamefont
  {Cowan}, \citenamefont {Roth}, \citenamefont {Phillips}, \citenamefont
  {Stoyer}, \citenamefont {Henry}, \citenamefont {Sangster}, \citenamefont
  {Singh}, \citenamefont {Wilks}, \citenamefont {MacKinnon}, \citenamefont
  {Offenberger}, \citenamefont {Pennington}, \citenamefont {Yasuike},
  \citenamefont {Langdon}, \citenamefont {Lasinski}, \citenamefont {Johnson},
  \citenamefont {Perry},\ and\ \citenamefont {Campbell}}]{skh00}%
  \BibitemOpen
  \bibfield  {author} {\bibinfo {author} {\bibfnamefont {R.~A.}\ \bibnamefont
  {Snavely}}, \bibinfo {author} {\bibfnamefont {M.~H.}\ \bibnamefont {Key}},
  \bibinfo {author} {\bibfnamefont {S.~P.}\ \bibnamefont {Hatchett}}, \bibinfo
  {author} {\bibfnamefont {T.~E.}\ \bibnamefont {Cowan}}, \bibinfo {author}
  {\bibfnamefont {M.}~\bibnamefont {Roth}}, \bibinfo {author} {\bibfnamefont
  {T.~W.}\ \bibnamefont {Phillips}}, \bibinfo {author} {\bibfnamefont {M.~A.}\
  \bibnamefont {Stoyer}}, \bibinfo {author} {\bibfnamefont {E.~A.}\
  \bibnamefont {Henry}}, \bibinfo {author} {\bibfnamefont {T.~C.}\ \bibnamefont
  {Sangster}}, \bibinfo {author} {\bibfnamefont {M.~S.}\ \bibnamefont {Singh}},
  \bibinfo {author} {\bibfnamefont {S.~C.}\ \bibnamefont {Wilks}}, \bibinfo
  {author} {\bibfnamefont {A.}~\bibnamefont {MacKinnon}}, \bibinfo {author}
  {\bibfnamefont {A.}~\bibnamefont {Offenberger}}, \bibinfo {author}
  {\bibfnamefont {D.~M.}\ \bibnamefont {Pennington}}, \bibinfo {author}
  {\bibfnamefont {K.}~\bibnamefont {Yasuike}}, \bibinfo {author} {\bibfnamefont
  {A.~B.}\ \bibnamefont {Langdon}}, \bibinfo {author} {\bibfnamefont {B.~F.}\
  \bibnamefont {Lasinski}}, \bibinfo {author} {\bibfnamefont {J.}~\bibnamefont
  {Johnson}}, \bibinfo {author} {\bibfnamefont {M.~D.}\ \bibnamefont {Perry}},
  \ and\ \bibinfo {author} {\bibfnamefont {E.~M.}\ \bibnamefont {Campbell}},\
  }\href {\doibase 10.1103/PhysRevLett.85.2945} {\bibfield  {journal} {\bibinfo
   {journal} {Phys. Rev. Lett.}\ }\textbf {\bibinfo {volume} {85}},\ \bibinfo
  {pages} {2945} (\bibinfo {year} {2000})}\BibitemShut {NoStop}%
\bibitem [{\citenamefont {Maksimchuk}\ \emph {et~al.}(2000)\citenamefont
  {Maksimchuk}, \citenamefont {Gu}, \citenamefont {Flippo}, \citenamefont
  {Umstadter},\ and\ \citenamefont {Bychenkov}}]{mgf00}%
  \BibitemOpen
  \bibfield  {author} {\bibinfo {author} {\bibfnamefont {A.}~\bibnamefont
  {Maksimchuk}}, \bibinfo {author} {\bibfnamefont {S.}~\bibnamefont {Gu}},
  \bibinfo {author} {\bibfnamefont {K.}~\bibnamefont {Flippo}}, \bibinfo
  {author} {\bibfnamefont {D.}~\bibnamefont {Umstadter}}, \ and\ \bibinfo
  {author} {\bibfnamefont {V.~Y.}\ \bibnamefont {Bychenkov}},\ }\href@noop {}
  {\bibfield  {journal} {\bibinfo  {journal} {Phys. Rev. Lett.}\ }\textbf
  {\bibinfo {volume} {84}},\ \bibinfo {pages} {4108} (\bibinfo {year}
  {2000})}\BibitemShut {NoStop}%
\bibitem [{\citenamefont {Badziak}\ \emph {et~al.}(2001)\citenamefont
  {Badziak}, \citenamefont {Woryna}, \citenamefont {Parys}, \citenamefont
  {Platonov}, \citenamefont {Jab\l{}o\ifmmode~\acute{n}\else \'{n}\fi{}ski},
  \citenamefont {Ry\ifmmode~\acute{c}\else \'{c}\fi{}}, \citenamefont
  {Vankov},\ and\ \citenamefont {Wo\l{}owski}}]{bwp01}%
  \BibitemOpen
  \bibfield  {author} {\bibinfo {author} {\bibfnamefont {J.}~\bibnamefont
  {Badziak}}, \bibinfo {author} {\bibfnamefont {E.}~\bibnamefont {Woryna}},
  \bibinfo {author} {\bibfnamefont {P.}~\bibnamefont {Parys}}, \bibinfo
  {author} {\bibfnamefont {K.~Y.}\ \bibnamefont {Platonov}}, \bibinfo {author}
  {\bibfnamefont {S.}~\bibnamefont {Jab\l{}o\ifmmode~\acute{n}\else
  \'{n}\fi{}ski}}, \bibinfo {author} {\bibfnamefont {L.}~\bibnamefont
  {Ry\ifmmode~\acute{c}\else \'{c}\fi{}}}, \bibinfo {author} {\bibfnamefont
  {A.~B.}\ \bibnamefont {Vankov}}, \ and\ \bibinfo {author} {\bibfnamefont
  {J.}~\bibnamefont {Wo\l{}owski}},\ }\href@noop {} {\bibfield  {journal}
  {\bibinfo  {journal} {Phys. Rev. Lett.}\ }\textbf {\bibinfo {volume} {87}},\
  \bibinfo {pages} {215001} (\bibinfo {year} {2001})}\BibitemShut {NoStop}%
\bibitem [{\citenamefont {Fuchs}\ \emph {et~al.}(2005)\citenamefont {Fuchs},
  \citenamefont {Sentoku}, \citenamefont {Karsch}, \citenamefont {Cobble},
  \citenamefont {Audebert}, \citenamefont {Kemp}, \citenamefont {Nikroo},
  \citenamefont {Antici}, \citenamefont {Brambrink}, \citenamefont {Blazevic},
  \citenamefont {Campbell}, \citenamefont {Fernández}, \citenamefont
  {Gauthier}, \citenamefont {Geissel}, \citenamefont {Hegelich}, \citenamefont
  {Pépin}, \citenamefont {Popescu}, \citenamefont {Renard-LeGalloudec},
  \citenamefont {Roth}, \citenamefont {Schreiber}, \citenamefont {Stephens},\
  and\ \citenamefont {Cowan}}]{fsk05}%
  \BibitemOpen
  \bibfield  {author} {\bibinfo {author} {\bibfnamefont {J.}~\bibnamefont
  {Fuchs}}, \bibinfo {author} {\bibfnamefont {Y.}~\bibnamefont {Sentoku}},
  \bibinfo {author} {\bibfnamefont {S.}~\bibnamefont {Karsch}}, \bibinfo
  {author} {\bibfnamefont {J.}~\bibnamefont {Cobble}}, \bibinfo {author}
  {\bibfnamefont {P.}~\bibnamefont {Audebert}}, \bibinfo {author}
  {\bibfnamefont {A.}~\bibnamefont {Kemp}}, \bibinfo {author} {\bibfnamefont
  {A.}~\bibnamefont {Nikroo}}, \bibinfo {author} {\bibfnamefont
  {P.}~\bibnamefont {Antici}}, \bibinfo {author} {\bibfnamefont
  {E.}~\bibnamefont {Brambrink}}, \bibinfo {author} {\bibfnamefont
  {A.}~\bibnamefont {Blazevic}}, \bibinfo {author} {\bibfnamefont {E.~M.}\
  \bibnamefont {Campbell}}, \bibinfo {author} {\bibfnamefont {J.~C.}\
  \bibnamefont {Fernández}}, \bibinfo {author} {\bibfnamefont
  {J.}~\bibnamefont {Gauthier}}, \bibinfo {author} {\bibfnamefont
  {M.}~\bibnamefont {Geissel}}, \bibinfo {author} {\bibfnamefont
  {M.}~\bibnamefont {Hegelich}}, \bibinfo {author} {\bibfnamefont
  {H.}~\bibnamefont {Pépin}}, \bibinfo {author} {\bibfnamefont
  {H.}~\bibnamefont {Popescu}}, \bibinfo {author} {\bibfnamefont
  {N.}~\bibnamefont {Renard-LeGalloudec}}, \bibinfo {author} {\bibfnamefont
  {M.}~\bibnamefont {Roth}}, \bibinfo {author} {\bibfnamefont {J.}~\bibnamefont
  {Schreiber}}, \bibinfo {author} {\bibfnamefont {R.}~\bibnamefont {Stephens}},
  \ and\ \bibinfo {author} {\bibfnamefont {T.~E.}\ \bibnamefont {Cowan}},\
  }\href@noop {} {\bibfield  {journal} {\bibinfo  {journal} {{Phys. Rev.
  Lett.}}\ }\textbf {\bibinfo {volume} {94}},\ \bibinfo {pages} {045004}
  (\bibinfo {year} {2005})}\BibitemShut {NoStop}%
\bibitem [{\citenamefont {Romagnani}\ \emph {et~al.}(2005)\citenamefont
  {Romagnani}, \citenamefont {Fuchs}, \citenamefont {Borghesi}, \citenamefont
  {Antici}, \citenamefont {Audebert}, \citenamefont {Ceccherini}, \citenamefont
  {Cowan}, \citenamefont {Grismayer}, \citenamefont {Kar}, \citenamefont
  {Macchi}, \citenamefont {Mora}, \citenamefont {Pretzler}, \citenamefont
  {Schiavi}, \citenamefont {Toncian},\ and\ \citenamefont {Willi}}]{rfb05}%
  \BibitemOpen
  \bibfield  {author} {\bibinfo {author} {\bibfnamefont {L.}~\bibnamefont
  {Romagnani}}, \bibinfo {author} {\bibfnamefont {J.}~\bibnamefont {Fuchs}},
  \bibinfo {author} {\bibfnamefont {M.}~\bibnamefont {Borghesi}}, \bibinfo
  {author} {\bibfnamefont {P.}~\bibnamefont {Antici}}, \bibinfo {author}
  {\bibfnamefont {P.}~\bibnamefont {Audebert}}, \bibinfo {author}
  {\bibfnamefont {F.}~\bibnamefont {Ceccherini}}, \bibinfo {author}
  {\bibfnamefont {T.}~\bibnamefont {Cowan}}, \bibinfo {author} {\bibfnamefont
  {T.}~\bibnamefont {Grismayer}}, \bibinfo {author} {\bibfnamefont
  {S.}~\bibnamefont {Kar}}, \bibinfo {author} {\bibfnamefont {A.}~\bibnamefont
  {Macchi}}, \bibinfo {author} {\bibfnamefont {P.}~\bibnamefont {Mora}},
  \bibinfo {author} {\bibfnamefont {G.}~\bibnamefont {Pretzler}}, \bibinfo
  {author} {\bibfnamefont {A.}~\bibnamefont {Schiavi}}, \bibinfo {author}
  {\bibfnamefont {T.}~\bibnamefont {Toncian}}, \ and\ \bibinfo {author}
  {\bibfnamefont {O.}~\bibnamefont {Willi}},\ }\href@noop {} {\bibfield
  {journal} {\bibinfo  {journal} {{Phys. Rev. Lett.}}\ }\textbf {\bibinfo
  {volume} {95}},\ \bibinfo {pages} {195001} (\bibinfo {year}
  {2005})}\BibitemShut {NoStop}%
\bibitem [{\citenamefont {Murakami}\ \emph {et~al.}(2005)\citenamefont
  {Murakami}, \citenamefont {Kang}, \citenamefont {Nishihara}, \citenamefont
  {Fujioka},\ and\ \citenamefont {Nishimura}}]{mkn05}%
  \BibitemOpen
  \bibfield  {author} {\bibinfo {author} {\bibfnamefont {M.}~\bibnamefont
  {Murakami}}, \bibinfo {author} {\bibfnamefont {Y.-G.}\ \bibnamefont {Kang}},
  \bibinfo {author} {\bibfnamefont {K.}~\bibnamefont {Nishihara}}, \bibinfo
  {author} {\bibfnamefont {S.}~\bibnamefont {Fujioka}}, \ and\ \bibinfo
  {author} {\bibfnamefont {H.}~\bibnamefont {Nishimura}},\ }\href@noop {}
  {\bibfield  {journal} {\bibinfo  {journal} {Phys. Plasmas}\ }\textbf
  {\bibinfo {volume} {12}},\ \bibinfo {eid} {062706} (\bibinfo {year}
  {2005})}\BibitemShut {NoStop}%
\bibitem [{\citenamefont {Mora}(2005)}]{mor05physplasmas}%
  \BibitemOpen
  \bibfield  {author} {\bibinfo {author} {\bibfnamefont {P.}~\bibnamefont
  {Mora}},\ }\href@noop {} {\bibfield  {journal} {\bibinfo  {journal} {Phys.
  Plasmas}\ }\textbf {\bibinfo {volume} {12}},\ \bibinfo {eid} {112102}
  (\bibinfo {year} {2005})}\BibitemShut {NoStop}%
\bibitem [{\citenamefont {Grismayer}\ and\ \citenamefont {Mora}(2006)}]{gmo06}%
  \BibitemOpen
  \bibfield  {author} {\bibinfo {author} {\bibfnamefont {T.}~\bibnamefont
  {Grismayer}}\ and\ \bibinfo {author} {\bibfnamefont {P.}~\bibnamefont
  {Mora}},\ }\href@noop {} {\bibfield  {journal} {\bibinfo  {journal} {Phys.
  Plasmas}\ }\textbf {\bibinfo {volume} {13}},\ \bibinfo {eid} {032103}
  (\bibinfo {year} {2006})}\BibitemShut {NoStop}%
\bibitem [{\citenamefont {Symes}\ \emph {et~al.}(2007)\citenamefont {Symes},
  \citenamefont {Hohenberger}, \citenamefont {Henig},\ and\ \citenamefont
  {Ditmire}}]{shh07}%
  \BibitemOpen
  \bibfield  {author} {\bibinfo {author} {\bibfnamefont {D.~R.}\ \bibnamefont
  {Symes}}, \bibinfo {author} {\bibfnamefont {M.}~\bibnamefont {Hohenberger}},
  \bibinfo {author} {\bibfnamefont {A.}~\bibnamefont {Henig}}, \ and\ \bibinfo
  {author} {\bibfnamefont {T.}~\bibnamefont {Ditmire}},\ }\href@noop {}
  {\bibfield  {journal} {\bibinfo  {journal} {Phys. Rev. Lett.}\ }\textbf
  {\bibinfo {volume} {98}},\ \bibinfo {eid} {123401} (\bibinfo {year}
  {2007})}\BibitemShut {NoStop}%
\bibitem [{\citenamefont {Kulin}\ \emph {et~al.}(2000)\citenamefont {Kulin},
  \citenamefont {Killian}, \citenamefont {Bergeson},\ and\ \citenamefont
  {Rolston}}]{kkb00}%
  \BibitemOpen
  \bibfield  {author} {\bibinfo {author} {\bibfnamefont {S.}~\bibnamefont
  {Kulin}}, \bibinfo {author} {\bibfnamefont {T.~C.}\ \bibnamefont {Killian}},
  \bibinfo {author} {\bibfnamefont {S.~D.}\ \bibnamefont {Bergeson}}, \ and\
  \bibinfo {author} {\bibfnamefont {S.~L.}\ \bibnamefont {Rolston}},\
  }\href@noop {} {\bibfield  {journal} {\bibinfo  {journal} {{Phys. Rev.
  Lett.}}\ }\textbf {\bibinfo {volume} {85}},\ \bibinfo {pages} {318} (\bibinfo
  {year} {2000})}\BibitemShut {NoStop}%
\bibitem [{\citenamefont {Robicheaux}\ and\ \citenamefont
  {Hanson}(2002)}]{rha02}%
  \BibitemOpen
  \bibfield  {author} {\bibinfo {author} {\bibfnamefont {F.}~\bibnamefont
  {Robicheaux}}\ and\ \bibinfo {author} {\bibfnamefont {J.~D.}\ \bibnamefont
  {Hanson}},\ }\href@noop {} {\bibfield  {journal} {\bibinfo  {journal} {{Phys.
  Rev. Lett.}}\ }\textbf {\bibinfo {volume} {88}},\ \bibinfo {pages} {55002}
  (\bibinfo {year} {2002})}\BibitemShut {NoStop}%
\bibitem [{\citenamefont {Robicheaux}\ and\ \citenamefont
  {Hanson}(2003)}]{rha03}%
  \BibitemOpen
  \bibfield  {author} {\bibinfo {author} {\bibfnamefont {F.}~\bibnamefont
  {Robicheaux}}\ and\ \bibinfo {author} {\bibfnamefont {J.~D.}\ \bibnamefont
  {Hanson}},\ }\href@noop {} {\bibfield  {journal} {\bibinfo  {journal} {{
  Phys. Plasmas}}\ }\textbf {\bibinfo {volume} {10}},\ \bibinfo {pages} {2217}
  (\bibinfo {year} {2003})}\BibitemShut {NoStop}%
\bibitem [{\citenamefont {Laha}\ \emph {et~al.}(2007)\citenamefont {Laha},
  \citenamefont {Gupta}, \citenamefont {Simien}, \citenamefont {Gao},
  \citenamefont {Castro},\ and\ \citenamefont {Killian}}]{lgs07}%
  \BibitemOpen
  \bibfield  {author} {\bibinfo {author} {\bibfnamefont {S.}~\bibnamefont
  {Laha}}, \bibinfo {author} {\bibfnamefont {P.}~\bibnamefont {Gupta}},
  \bibinfo {author} {\bibfnamefont {C.~E.}\ \bibnamefont {Simien}}, \bibinfo
  {author} {\bibfnamefont {H.}~\bibnamefont {Gao}}, \bibinfo {author}
  {\bibfnamefont {J.}~\bibnamefont {Castro}}, \ and\ \bibinfo {author}
  {\bibfnamefont {T.~C.}\ \bibnamefont {Killian}},\ }\href@noop {} {\bibfield
  {journal} {\bibinfo  {journal} {{Phys. Rev. Lett.}}\ }\textbf {\bibinfo
  {volume} {99}},\ \bibinfo {pages} {155001} (\bibinfo {year}
  {2007})}\BibitemShut {NoStop}%
\bibitem [{\citenamefont {Gorman}\ \emph
  {et~al.}(2020{\natexlab{a}})\citenamefont {Gorman}, \citenamefont {Warrens},
  \citenamefont {Bradshaw},\ and\ \citenamefont {Killian}}]{gwb20}%
  \BibitemOpen
  \bibfield  {author} {\bibinfo {author} {\bibfnamefont {G.~M.}\ \bibnamefont
  {Gorman}}, \bibinfo {author} {\bibfnamefont {M.~K.}\ \bibnamefont {Warrens}},
  \bibinfo {author} {\bibfnamefont {S.~J.}\ \bibnamefont {Bradshaw}}, \ and\
  \bibinfo {author} {\bibfnamefont {T.~C.}\ \bibnamefont {Killian}},\
  }\href@noop {} {} (\bibinfo {year} {2020}{\natexlab{a}}),\ \Eprint
  {http://arxiv.org/abs/2010.16355} {arXiv:2010.16355 [physics.plasm-ph]}
  \BibitemShut {NoStop}%
\bibitem [{\citenamefont {Killian}\ \emph {et~al.}(1999)\citenamefont
  {Killian}, \citenamefont {Kulin}, \citenamefont {Bergeson}, \citenamefont
  {Orozco}, \citenamefont {Orzel},\ and\ \citenamefont {Rolston}}]{kkb99}%
  \BibitemOpen
  \bibfield  {author} {\bibinfo {author} {\bibfnamefont {T.~C.}\ \bibnamefont
  {Killian}}, \bibinfo {author} {\bibfnamefont {S.}~\bibnamefont {Kulin}},
  \bibinfo {author} {\bibfnamefont {S.~D.}\ \bibnamefont {Bergeson}}, \bibinfo
  {author} {\bibfnamefont {L.~A.}\ \bibnamefont {Orozco}}, \bibinfo {author}
  {\bibfnamefont {C.}~\bibnamefont {Orzel}}, \ and\ \bibinfo {author}
  {\bibfnamefont {S.~L.}\ \bibnamefont {Rolston}},\ }\href@noop {} {\bibfield
  {journal} {\bibinfo  {journal} {{Phys. Rev. Lett.}}\ }\textbf {\bibinfo
  {volume} {83}},\ \bibinfo {pages} {4776} (\bibinfo {year}
  {1999})}\BibitemShut {NoStop}%
\bibitem [{\citenamefont {Killian}\ \emph {et~al.}(2007)\citenamefont
  {Killian}, \citenamefont {Pattard}, \citenamefont {Pohl},\ and\ \citenamefont
  {Rost}}]{kpp07}%
  \BibitemOpen
  \bibfield  {author} {\bibinfo {author} {\bibfnamefont {T.~C.}\ \bibnamefont
  {Killian}}, \bibinfo {author} {\bibfnamefont {T.}~\bibnamefont {Pattard}},
  \bibinfo {author} {\bibfnamefont {T.}~\bibnamefont {Pohl}}, \ and\ \bibinfo
  {author} {\bibfnamefont {J.~M.}\ \bibnamefont {Rost}},\ }\href@noop {}
  {\bibfield  {journal} {\bibinfo  {journal} {{Phys. Rep.}}\ }\textbf {\bibinfo
  {volume} {449}},\ \bibinfo {pages} {77} (\bibinfo {year} {2007})}\BibitemShut
  {NoStop}%
\bibitem [{\citenamefont {Lyon}\ and\ \citenamefont {Rolston}(2017)}]{lro17}%
  \BibitemOpen
  \bibfield  {author} {\bibinfo {author} {\bibfnamefont {M.}~\bibnamefont
  {Lyon}}\ and\ \bibinfo {author} {\bibfnamefont {S.~L.}\ \bibnamefont
  {Rolston}},\ }\href@noop {} {\bibfield  {journal} {\bibinfo  {journal} {Rep.
  Prog. Phys.}\ }\textbf {\bibinfo {volume} {80}},\ \bibinfo {pages} {017001}
  (\bibinfo {year} {2017})}\BibitemShut {NoStop}%
\bibitem [{\citenamefont {Metcalf}\ and\ \citenamefont {van~der
  Straten}(1999)}]{mvs99}%
  \BibitemOpen
  \bibfield  {author} {\bibinfo {author} {\bibfnamefont {H.~J.}\ \bibnamefont
  {Metcalf}}\ and\ \bibinfo {author} {\bibfnamefont {P.}~\bibnamefont {van~der
  Straten}},\ }\href@noop {} {\emph {\bibinfo {title} {Laser Cooling and
  Trapping}}}\ (\bibinfo  {publisher} {Springer-Verlag},\ \bibinfo {address}
  {New York, New York},\ \bibinfo {year} {1999})\BibitemShut {NoStop}%
\bibitem [{\citenamefont {Morrison}\ \emph {et~al.}(2008)\citenamefont
  {Morrison}, \citenamefont {Rennick}, \citenamefont {Keller},\ and\
  \citenamefont {Grant}}]{mrk08}%
  \BibitemOpen
  \bibfield  {author} {\bibinfo {author} {\bibfnamefont {J.~P.}\ \bibnamefont
  {Morrison}}, \bibinfo {author} {\bibfnamefont {C.~J.}\ \bibnamefont
  {Rennick}}, \bibinfo {author} {\bibfnamefont {J.~S.}\ \bibnamefont {Keller}},
  \ and\ \bibinfo {author} {\bibfnamefont {E.~R.}\ \bibnamefont {Grant}},\
  }\href@noop {} {\bibfield  {journal} {\bibinfo  {journal} {Phys. Rev. Lett.}\
  }\textbf {\bibinfo {volume} {101}},\ \bibinfo {eid} {205005} (\bibinfo {year}
  {2008})}\BibitemShut {NoStop}%
\bibitem [{\citenamefont {Simien}\ \emph {et~al.}(2004)\citenamefont {Simien},
  \citenamefont {Chen}, \citenamefont {Gupta}, \citenamefont {Laha},
  \citenamefont {Martinez}, \citenamefont {Mickelson}, \citenamefont {Nagel},\
  and\ \citenamefont {Killian}}]{scg04}%
  \BibitemOpen
  \bibfield  {author} {\bibinfo {author} {\bibfnamefont {C.~E.}\ \bibnamefont
  {Simien}}, \bibinfo {author} {\bibfnamefont {Y.~C.}\ \bibnamefont {Chen}},
  \bibinfo {author} {\bibfnamefont {P.}~\bibnamefont {Gupta}}, \bibinfo
  {author} {\bibfnamefont {S.}~\bibnamefont {Laha}}, \bibinfo {author}
  {\bibfnamefont {Y.~N.}\ \bibnamefont {Martinez}}, \bibinfo {author}
  {\bibfnamefont {P.~G.}\ \bibnamefont {Mickelson}}, \bibinfo {author}
  {\bibfnamefont {S.~B.}\ \bibnamefont {Nagel}}, \ and\ \bibinfo {author}
  {\bibfnamefont {T.~C.}\ \bibnamefont {Killian}},\ }\href@noop {} {\bibfield
  {journal} {\bibinfo  {journal} {{Phys. Rev. Lett.}}\ }\textbf {\bibinfo
  {volume} {92}},\ \bibinfo {pages} {143001} (\bibinfo {year}
  {2004})}\BibitemShut {NoStop}%
\bibitem [{\citenamefont {Cornolti}\ \emph {et~al.}(2005)\citenamefont
  {Cornolti}, \citenamefont {Ceccherini}, \citenamefont {Betti},\ and\
  \citenamefont {Pegoraro}}]{ccb05}%
  \BibitemOpen
  \bibfield  {author} {\bibinfo {author} {\bibfnamefont {F.}~\bibnamefont
  {Cornolti}}, \bibinfo {author} {\bibfnamefont {F.}~\bibnamefont
  {Ceccherini}}, \bibinfo {author} {\bibfnamefont {S.}~\bibnamefont {Betti}}, \
  and\ \bibinfo {author} {\bibfnamefont {F.}~\bibnamefont {Pegoraro}},\
  }\href@noop {} {\bibfield  {journal} {\bibinfo  {journal} {Phys. Rev. E}\
  }\textbf {\bibinfo {volume} {71}},\ \bibinfo {eid} {056407} (\bibinfo {year}
  {2005})}\BibitemShut {NoStop}%
\bibitem [{\citenamefont {Murillo}(2001)}]{mur01}%
  \BibitemOpen
  \bibfield  {author} {\bibinfo {author} {\bibfnamefont {M.~S.}\ \bibnamefont
  {Murillo}},\ }\href@noop {} {\bibfield  {journal} {\bibinfo  {journal}
  {{Phys. Rev. Lett.}}\ }\textbf {\bibinfo {volume} {87}},\ \bibinfo {pages}
  {115003} (\bibinfo {year} {2001})}\BibitemShut {NoStop}%
\bibitem [{\citenamefont {Kuzmin}\ and\ \citenamefont {O'Neil}(2002)}]{kon02}%
  \BibitemOpen
  \bibfield  {author} {\bibinfo {author} {\bibfnamefont {S.~G.}\ \bibnamefont
  {Kuzmin}}\ and\ \bibinfo {author} {\bibfnamefont {T.~M.}\ \bibnamefont
  {O'Neil}},\ }\href@noop {} {\bibfield  {journal} {\bibinfo  {journal} {{Phys.
  Plasmas}}\ }\textbf {\bibinfo {volume} {9}},\ \bibinfo {pages} {3743}
  (\bibinfo {year} {2002})}\BibitemShut {NoStop}%
\bibitem [{\citenamefont {Mazevet}, \citenamefont {Collins},\ and\
  \citenamefont {Kress}(2002)}]{mck02}%
  \BibitemOpen
  \bibfield  {author} {\bibinfo {author} {\bibfnamefont {S.}~\bibnamefont
  {Mazevet}}, \bibinfo {author} {\bibfnamefont {L.~A.}\ \bibnamefont
  {Collins}}, \ and\ \bibinfo {author} {\bibfnamefont {J.~D.}\ \bibnamefont
  {Kress}},\ }\href@noop {} {\bibfield  {journal} {\bibinfo  {journal} {{Phys.
  Rev. Lett.}}\ }\textbf {\bibinfo {volume} {88}},\ \bibinfo {pages} {55001}
  (\bibinfo {year} {2002})}\BibitemShut {NoStop}%
\bibitem [{\citenamefont {Gericke}\ \emph {et~al.}(2003)\citenamefont
  {Gericke}, \citenamefont {Murillo}, \citenamefont {Semkat}, \citenamefont
  {Bonitz},\ and\ \citenamefont {Kremp}}]{gms03}%
  \BibitemOpen
  \bibfield  {author} {\bibinfo {author} {\bibfnamefont {D.~O.}\ \bibnamefont
  {Gericke}}, \bibinfo {author} {\bibfnamefont {M.~S.}\ \bibnamefont
  {Murillo}}, \bibinfo {author} {\bibfnamefont {D.}~\bibnamefont {Semkat}},
  \bibinfo {author} {\bibfnamefont {M.}~\bibnamefont {Bonitz}}, \ and\ \bibinfo
  {author} {\bibfnamefont {D.}~\bibnamefont {Kremp}},\ }\href@noop {}
  {\bibfield  {journal} {\bibinfo  {journal} {{J. Phys. A}}\ }\textbf {\bibinfo
  {volume} {36}},\ \bibinfo {pages} {6087} (\bibinfo {year}
  {2003})}\BibitemShut {NoStop}%
\bibitem [{\citenamefont {Murillo}(2006)}]{mur06PRL}%
  \BibitemOpen
  \bibfield  {author} {\bibinfo {author} {\bibfnamefont {M.~S.}\ \bibnamefont
  {Murillo}},\ }\href@noop {} {\bibfield  {journal} {\bibinfo  {journal}
  {{Phys. Rev. Lett.}}\ }\textbf {\bibinfo {volume} {96}},\ \bibinfo {pages}
  {165001} (\bibinfo {year} {2006})}\BibitemShut {NoStop}%
\bibitem [{\citenamefont {Murillo}(2007)}]{mur07PoP}%
  \BibitemOpen
  \bibfield  {author} {\bibinfo {author} {\bibfnamefont {M.~S.}\ \bibnamefont
  {Murillo}},\ }\href {\doibase 10.1063/1.2436853} {\bibfield  {journal}
  {\bibinfo  {journal} {Phys. Plasmas}\ }\textbf {\bibinfo {volume} {14}},\
  \bibinfo {pages} {055702} (\bibinfo {year} {2007})}\BibitemShut {NoStop}%
\bibitem [{\citenamefont {Pohl}, \citenamefont {Pattard},\ and\ \citenamefont
  {Rost}(2004)}]{ppr04PRA}%
  \BibitemOpen
  \bibfield  {author} {\bibinfo {author} {\bibfnamefont {T.}~\bibnamefont
  {Pohl}}, \bibinfo {author} {\bibfnamefont {T.}~\bibnamefont {Pattard}}, \
  and\ \bibinfo {author} {\bibfnamefont {J.~M.}\ \bibnamefont {Rost}},\
  }\href@noop {} {\bibfield  {journal} {\bibinfo  {journal} {{ Phys. Rev. A}}\
  }\textbf {\bibinfo {volume} {70}},\ \bibinfo {pages} {033416} (\bibinfo
  {year} {2004})}\BibitemShut {NoStop}%
\bibitem [{\citenamefont {Pohl}, \citenamefont {Pattard},\ and\ \citenamefont
  {Rost}(2005)}]{ppr05PRL}%
  \BibitemOpen
  \bibfield  {author} {\bibinfo {author} {\bibfnamefont {T.}~\bibnamefont
  {Pohl}}, \bibinfo {author} {\bibfnamefont {T.}~\bibnamefont {Pattard}}, \
  and\ \bibinfo {author} {\bibfnamefont {J.~M.}\ \bibnamefont {Rost}},\
  }\href@noop {} {\bibfield  {journal} {\bibinfo  {journal} {{ Phys. Rev.
  Lett.}}\ }\textbf {\bibinfo {volume} {94}},\ \bibinfo {pages} {205003}
  (\bibinfo {year} {2005})}\BibitemShut {NoStop}%
\bibitem [{\citenamefont {Chen}\ \emph {et~al.}(2004)\citenamefont {Chen},
  \citenamefont {Simien}, \citenamefont {Laha}, \citenamefont {Gupta},
  \citenamefont {Martinez}, \citenamefont {Mickelson}, \citenamefont {Nagel},\
  and\ \citenamefont {Killian}}]{csl04}%
  \BibitemOpen
  \bibfield  {author} {\bibinfo {author} {\bibfnamefont {Y.~C.}\ \bibnamefont
  {Chen}}, \bibinfo {author} {\bibfnamefont {C.~E.}\ \bibnamefont {Simien}},
  \bibinfo {author} {\bibfnamefont {S.}~\bibnamefont {Laha}}, \bibinfo {author}
  {\bibfnamefont {P.}~\bibnamefont {Gupta}}, \bibinfo {author} {\bibfnamefont
  {Y.~N.}\ \bibnamefont {Martinez}}, \bibinfo {author} {\bibfnamefont {P.~G.}\
  \bibnamefont {Mickelson}}, \bibinfo {author} {\bibfnamefont {S.~B.}\
  \bibnamefont {Nagel}}, \ and\ \bibinfo {author} {\bibfnamefont {T.~C.}\
  \bibnamefont {Killian}},\ }\href@noop {} {\bibfield  {journal} {\bibinfo
  {journal} {{Phys. Rev. Lett.}}\ }\textbf {\bibinfo {volume} {93}},\ \bibinfo
  {pages} {265003} (\bibinfo {year} {2004})}\BibitemShut {NoStop}%
\bibitem [{\citenamefont {Laha}\ \emph {et~al.}(2006)\citenamefont {Laha},
  \citenamefont {Chen}, \citenamefont {Gupta}, \citenamefont {Simien},
  \citenamefont {Martinez}, \citenamefont {Mickelson}, \citenamefont {Nagel},\
  and\ \citenamefont {Killian}}]{lcg06}%
  \BibitemOpen
  \bibfield  {author} {\bibinfo {author} {\bibfnamefont {S.}~\bibnamefont
  {Laha}}, \bibinfo {author} {\bibfnamefont {Y.~C.}\ \bibnamefont {Chen}},
  \bibinfo {author} {\bibfnamefont {P.}~\bibnamefont {Gupta}}, \bibinfo
  {author} {\bibfnamefont {C.~E.}\ \bibnamefont {Simien}}, \bibinfo {author}
  {\bibfnamefont {Y.~N.}\ \bibnamefont {Martinez}}, \bibinfo {author}
  {\bibfnamefont {P.~G.}\ \bibnamefont {Mickelson}}, \bibinfo {author}
  {\bibfnamefont {S.~B.}\ \bibnamefont {Nagel}}, \ and\ \bibinfo {author}
  {\bibfnamefont {T.~C.}\ \bibnamefont {Killian}},\ }\href@noop {} {\bibfield
  {journal} {\bibinfo  {journal} {{Eur. Phys. J. D}}\ }\textbf {\bibinfo
  {volume} {40}},\ \bibinfo {pages} {51} (\bibinfo {year} {2006})}\BibitemShut
  {NoStop}%
\bibitem [{\citenamefont {Bannasch}\ \emph {et~al.}(2012)\citenamefont
  {Bannasch}, \citenamefont {Castro}, \citenamefont {McQuillen}, \citenamefont
  {Pohl},\ and\ \citenamefont {Killian}}]{bcm12}%
  \BibitemOpen
  \bibfield  {author} {\bibinfo {author} {\bibfnamefont {G.}~\bibnamefont
  {Bannasch}}, \bibinfo {author} {\bibfnamefont {J.}~\bibnamefont {Castro}},
  \bibinfo {author} {\bibfnamefont {P.}~\bibnamefont {McQuillen}}, \bibinfo
  {author} {\bibfnamefont {T.}~\bibnamefont {Pohl}}, \ and\ \bibinfo {author}
  {\bibfnamefont {T.~C.}\ \bibnamefont {Killian}},\ }\href {\doibase
  10.1103/PhysRevLett.109.185008} {\bibfield  {journal} {\bibinfo  {journal}
  {Phys. Rev. Lett.}\ }\textbf {\bibinfo {volume} {109}},\ \bibinfo {pages}
  {185008} (\bibinfo {year} {2012})}\BibitemShut {NoStop}%
\bibitem [{\citenamefont {Lyon}, \citenamefont {Bergeson},\ and\ \citenamefont
  {Murillo}(2013)}]{lbm13}%
  \BibitemOpen
  \bibfield  {author} {\bibinfo {author} {\bibfnamefont {M.}~\bibnamefont
  {Lyon}}, \bibinfo {author} {\bibfnamefont {S.~D.}\ \bibnamefont {Bergeson}},
  \ and\ \bibinfo {author} {\bibfnamefont {M.~S.}\ \bibnamefont {Murillo}},\
  }\href {\doibase 10.1103/PhysRevE.87.033101} {\bibfield  {journal} {\bibinfo
  {journal} {Phys. Rev. E}\ }\textbf {\bibinfo {volume} {87}},\ \bibinfo
  {pages} {033101} (\bibinfo {year} {2013})}\BibitemShut {NoStop}%
\bibitem [{\citenamefont {Strickler}\ \emph {et~al.}(2016)\citenamefont
  {Strickler}, \citenamefont {Langin}, \citenamefont {McQuillen}, \citenamefont
  {Daligault},\ and\ \citenamefont {Killian}}]{slm16}%
  \BibitemOpen
  \bibfield  {author} {\bibinfo {author} {\bibfnamefont {T.~S.}\ \bibnamefont
  {Strickler}}, \bibinfo {author} {\bibfnamefont {T.~K.}\ \bibnamefont
  {Langin}}, \bibinfo {author} {\bibfnamefont {P.}~\bibnamefont {McQuillen}},
  \bibinfo {author} {\bibfnamefont {J.}~\bibnamefont {Daligault}}, \ and\
  \bibinfo {author} {\bibfnamefont {T.~C.}\ \bibnamefont {Killian}},\
  }\href@noop {} {\bibfield  {journal} {\bibinfo  {journal} {Phys. Rev. X}\
  }\textbf {\bibinfo {volume} {6}},\ \bibinfo {pages} {021021} (\bibinfo {year}
  {2016})}\BibitemShut {NoStop}%
\bibitem [{\citenamefont {Ichimaru}(1982)}]{ich82}%
  \BibitemOpen
  \bibfield  {author} {\bibinfo {author} {\bibfnamefont {S.}~\bibnamefont
  {Ichimaru}},\ }\href@noop {} {\bibfield  {journal} {\bibinfo  {journal}
  {{Rev. of Mod. Phys.}}\ }\textbf {\bibinfo {volume} {54}},\ \bibinfo {pages}
  {1017} (\bibinfo {year} {1982})}\BibitemShut {NoStop}%
\bibitem [{\citenamefont {Lyon}\ \emph {et~al.}(2015)\citenamefont {Lyon},
  \citenamefont {Bergeson}, \citenamefont {Hart},\ and\ \citenamefont
  {Murillo}}]{lbh15}%
  \BibitemOpen
  \bibfield  {author} {\bibinfo {author} {\bibfnamefont {M.}~\bibnamefont
  {Lyon}}, \bibinfo {author} {\bibfnamefont {S.~D.}\ \bibnamefont {Bergeson}},
  \bibinfo {author} {\bibfnamefont {G.}~\bibnamefont {Hart}}, \ and\ \bibinfo
  {author} {\bibfnamefont {M.}~\bibnamefont {Murillo}},\ }\href@noop {}
  {\bibfield  {journal} {\bibinfo  {journal} {Sci. Rep.}\ }\textbf {\bibinfo
  {volume} {5}},\ \bibinfo {pages} {1} (\bibinfo {year} {2015})}\BibitemShut
  {NoStop}%
\bibitem [{\citenamefont {Murillo}(2004)}]{mur04}%
  \BibitemOpen
  \bibfield  {author} {\bibinfo {author} {\bibfnamefont {M.~S.}\ \bibnamefont
  {Murillo}},\ }\href@noop {} {\bibfield  {journal} {\bibinfo  {journal} {Phys.
  Plasmas}\ }\textbf {\bibinfo {volume} {11}},\ \bibinfo {pages} {2964}
  (\bibinfo {year} {2004})}\BibitemShut {NoStop}%
\bibitem [{\citenamefont {Bergeson}\ \emph {et~al.}(2019)\citenamefont
  {Bergeson}, \citenamefont {Baalrud}, \citenamefont {{Leland Ellison}},
  \citenamefont {Grant}, \citenamefont {Graziani}, \citenamefont {Killian},
  \citenamefont {Murillo}, \citenamefont {Roberts},\ and\ \citenamefont
  {Stanton}}]{bbl19}%
  \BibitemOpen
  \bibfield  {author} {\bibinfo {author} {\bibfnamefont {S.~D.}\ \bibnamefont
  {Bergeson}}, \bibinfo {author} {\bibfnamefont {S.~D.}\ \bibnamefont
  {Baalrud}}, \bibinfo {author} {\bibfnamefont {C.}~\bibnamefont {{Leland
  Ellison}}}, \bibinfo {author} {\bibfnamefont {E.}~\bibnamefont {Grant}},
  \bibinfo {author} {\bibfnamefont {F.~R.}\ \bibnamefont {Graziani}}, \bibinfo
  {author} {\bibfnamefont {T.~C.}\ \bibnamefont {Killian}}, \bibinfo {author}
  {\bibfnamefont {M.~S.}\ \bibnamefont {Murillo}}, \bibinfo {author}
  {\bibfnamefont {J.~L.}\ \bibnamefont {Roberts}}, \ and\ \bibinfo {author}
  {\bibfnamefont {L.~G.}\ \bibnamefont {Stanton}},\ }\href@noop {} {\bibfield
  {journal} {\bibinfo  {journal} {Phys. Plasmas}\ }\textbf {\bibinfo {volume}
  {26}},\ \bibinfo {pages} {100501} (\bibinfo {year} {2019})}\BibitemShut
  {NoStop}%
\bibitem [{\citenamefont {Gorman}\ \emph
  {et~al.}(2020{\natexlab{b}})\citenamefont {Gorman}, \citenamefont {Langin},
  \citenamefont {Warrens}, \citenamefont {Vrinceanu},\ and\ \citenamefont
  {Killian}}]{glw20}%
  \BibitemOpen
  \bibfield  {author} {\bibinfo {author} {\bibfnamefont {G.~M.}\ \bibnamefont
  {Gorman}}, \bibinfo {author} {\bibfnamefont {T.~K.}\ \bibnamefont {Langin}},
  \bibinfo {author} {\bibfnamefont {M.~K.}\ \bibnamefont {Warrens}}, \bibinfo
  {author} {\bibfnamefont {D.}~\bibnamefont {Vrinceanu}}, \ and\ \bibinfo
  {author} {\bibfnamefont {T.~C.}\ \bibnamefont {Killian}},\ }\href {\doibase
  10.1103/PhysRevA.101.012710} {\bibfield  {journal} {\bibinfo  {journal}
  {Phys. Rev. A}\ }\textbf {\bibinfo {volume} {101}},\ \bibinfo {pages}
  {012710} (\bibinfo {year} {2020}{\natexlab{b}})}\BibitemShut {NoStop}%
\bibitem [{\citenamefont {McQuillen}\ \emph {et~al.}(2013)\citenamefont
  {McQuillen}, \citenamefont {Castro}, \citenamefont {Strickler}, \citenamefont
  {Bradshaw},\ and\ \citenamefont {Killian}}]{mcs13}%
  \BibitemOpen
  \bibfield  {author} {\bibinfo {author} {\bibfnamefont {P.}~\bibnamefont
  {McQuillen}}, \bibinfo {author} {\bibfnamefont {J.}~\bibnamefont {Castro}},
  \bibinfo {author} {\bibfnamefont {T.}~\bibnamefont {Strickler}}, \bibinfo
  {author} {\bibfnamefont {S.~J.}\ \bibnamefont {Bradshaw}}, \ and\ \bibinfo
  {author} {\bibfnamefont {T.~C.}\ \bibnamefont {Killian}},\ }\href@noop {}
  {\bibfield  {journal} {\bibinfo  {journal} {Phys. Plasmas}\ }\textbf
  {\bibinfo {volume} {20}},\ \bibinfo {pages} {043516} (\bibinfo {year}
  {2013})}\BibitemShut {NoStop}%
\bibitem [{\citenamefont {McQuillen}\ \emph
  {et~al.}(2015{\natexlab{a}})\citenamefont {McQuillen}, \citenamefont
  {Castro}, \citenamefont {Bradshaw},\ and\ \citenamefont {Killian}}]{mcb15}%
  \BibitemOpen
  \bibfield  {author} {\bibinfo {author} {\bibfnamefont {P.}~\bibnamefont
  {McQuillen}}, \bibinfo {author} {\bibfnamefont {J.}~\bibnamefont {Castro}},
  \bibinfo {author} {\bibfnamefont {S.~J.}\ \bibnamefont {Bradshaw}}, \ and\
  \bibinfo {author} {\bibfnamefont {T.~C.}\ \bibnamefont {Killian}},\
  }\href@noop {} {\bibfield  {journal} {\bibinfo  {journal} {Phys. of Plasmas}\
  }\textbf {\bibinfo {volume} {22}},\ \bibinfo {pages} {043514} (\bibinfo
  {year} {2015}{\natexlab{a}})}\BibitemShut {NoStop}%
\bibitem [{\citenamefont {Fletcher}, \citenamefont {Zhang},\ and\ \citenamefont
  {Rolston}(2006)}]{fzr06}%
  \BibitemOpen
  \bibfield  {author} {\bibinfo {author} {\bibfnamefont {R.~S.}\ \bibnamefont
  {Fletcher}}, \bibinfo {author} {\bibfnamefont {X.~L.}\ \bibnamefont {Zhang}},
  \ and\ \bibinfo {author} {\bibfnamefont {S.~L.}\ \bibnamefont {Rolston}},\
  }\href@noop {} {\bibfield  {journal} {\bibinfo  {journal} {{ Phys. Rev.
  Lett.}}\ }\textbf {\bibinfo {volume} {96}},\ \bibinfo {pages} {105003}
  (\bibinfo {year} {2006})}\BibitemShut {NoStop}%
\bibitem [{\citenamefont {Zhang}, \citenamefont {Fletcher},\ and\ \citenamefont
  {Rolston}(2008)}]{zfr08}%
  \BibitemOpen
  \bibfield  {author} {\bibinfo {author} {\bibfnamefont {X.~L.}\ \bibnamefont
  {Zhang}}, \bibinfo {author} {\bibfnamefont {R.~S.}\ \bibnamefont {Fletcher}},
  \ and\ \bibinfo {author} {\bibfnamefont {S.~L.}\ \bibnamefont {Rolston}},\
  }\href@noop {} {\bibfield  {journal} {\bibinfo  {journal} {Phys. Rev. Lett.}\
  }\textbf {\bibinfo {volume} {101}},\ \bibinfo {pages} {195002} (\bibinfo
  {year} {2008})}\BibitemShut {NoStop}%
\bibitem [{\citenamefont {Castro}, \citenamefont {McQuillen},\ and\
  \citenamefont {Killian}(2010)}]{cmk10}%
  \BibitemOpen
  \bibfield  {author} {\bibinfo {author} {\bibfnamefont {J.}~\bibnamefont
  {Castro}}, \bibinfo {author} {\bibfnamefont {P.}~\bibnamefont {McQuillen}}, \
  and\ \bibinfo {author} {\bibfnamefont {T.~C.}\ \bibnamefont {Killian}},\
  }\href@noop {} {\bibfield  {journal} {\bibinfo  {journal} {{Phys. Rev.
  Lett.}}\ }\textbf {\bibinfo {volume} {105}},\ \bibinfo {pages} {065004}
  (\bibinfo {year} {2010})}\BibitemShut {NoStop}%
\bibitem [{\citenamefont {Claessens}\ \emph {et~al.}(2005)\citenamefont
  {Claessens}, \citenamefont {van~der Geer}, \citenamefont {Taban},
  \citenamefont {Vredenbregt},\ and\ \citenamefont {Luiten}}]{cgt05}%
  \BibitemOpen
  \bibfield  {author} {\bibinfo {author} {\bibfnamefont {B.~J.}\ \bibnamefont
  {Claessens}}, \bibinfo {author} {\bibfnamefont {S.~B.}\ \bibnamefont {van~der
  Geer}}, \bibinfo {author} {\bibfnamefont {G.}~\bibnamefont {Taban}}, \bibinfo
  {author} {\bibfnamefont {E.~J.~D.}\ \bibnamefont {Vredenbregt}}, \ and\
  \bibinfo {author} {\bibfnamefont {O.~J.}\ \bibnamefont {Luiten}},\
  }\href@noop {} {\bibfield  {journal} {\bibinfo  {journal} {Phys. Rev. Lett.}\
  }\textbf {\bibinfo {volume} {95}},\ \bibinfo {pages} {164801} (\bibinfo
  {year} {2005})}\BibitemShut {NoStop}%
\bibitem [{\citenamefont {Reijnders}\ \emph {et~al.}(2009)\citenamefont
  {Reijnders}, \citenamefont {van Kruisbergen}, \citenamefont {Taban},
  \citenamefont {van~der Geer}, \citenamefont {Mutsaers}, \citenamefont
  {Vredenbregt},\ and\ \citenamefont {Luiten}}]{rkt09}%
  \BibitemOpen
  \bibfield  {author} {\bibinfo {author} {\bibfnamefont {M.~P.}\ \bibnamefont
  {Reijnders}}, \bibinfo {author} {\bibfnamefont {P.~A.}\ \bibnamefont {van
  Kruisbergen}}, \bibinfo {author} {\bibfnamefont {G.}~\bibnamefont {Taban}},
  \bibinfo {author} {\bibfnamefont {S.~B.}\ \bibnamefont {van~der Geer}},
  \bibinfo {author} {\bibfnamefont {P.~H.~A.}\ \bibnamefont {Mutsaers}},
  \bibinfo {author} {\bibfnamefont {E.~J.~D.}\ \bibnamefont {Vredenbregt}}, \
  and\ \bibinfo {author} {\bibfnamefont {O.~J.}\ \bibnamefont {Luiten}},\
  }\href {\doibase 10.1103/PhysRevLett.102.034802} {\bibfield  {journal}
  {\bibinfo  {journal} {Phys. Rev. Lett.}\ }\textbf {\bibinfo {volume} {102}},\
  \bibinfo {pages} {034802} (\bibinfo {year} {2009})}\BibitemShut {NoStop}%
\bibitem [{\citenamefont {McCulloch}\ \emph {et~al.}(2011)\citenamefont
  {McCulloch}, \citenamefont {Sheludko}, \citenamefont {Saliba}, \citenamefont
  {Bell}, \citenamefont {Junker}, \citenamefont {Nugent},\ and\ \citenamefont
  {Scholten}}]{mss11}%
  \BibitemOpen
  \bibfield  {author} {\bibinfo {author} {\bibfnamefont {A.~J.}\ \bibnamefont
  {McCulloch}}, \bibinfo {author} {\bibfnamefont {D.~V.}\ \bibnamefont
  {Sheludko}}, \bibinfo {author} {\bibfnamefont {S.~D.}\ \bibnamefont
  {Saliba}}, \bibinfo {author} {\bibfnamefont {S.~C.}\ \bibnamefont {Bell}},
  \bibinfo {author} {\bibfnamefont {M.}~\bibnamefont {Junker}}, \bibinfo
  {author} {\bibfnamefont {K.~A.}\ \bibnamefont {Nugent}}, \ and\ \bibinfo
  {author} {\bibfnamefont {R.~E.}\ \bibnamefont {Scholten}},\ }\href@noop {}
  {\bibfield  {journal} {\bibinfo  {journal} {Nature Physics}\ }\textbf
  {\bibinfo {volume} {7}},\ \bibinfo {pages} {1} (\bibinfo {year}
  {2011})}\BibitemShut {NoStop}%
\bibitem [{\citenamefont {McClelland}\ \emph {et~al.}(2016)\citenamefont
  {McClelland}, \citenamefont {Steele}, \citenamefont {Knuffman}, \citenamefont
  {Twedt}, \citenamefont {Schwarzkopf},\ and\ \citenamefont {Wilson}}]{msk16}%
  \BibitemOpen
  \bibfield  {author} {\bibinfo {author} {\bibfnamefont {J.~J.}\ \bibnamefont
  {McClelland}}, \bibinfo {author} {\bibfnamefont {A.~V.}\ \bibnamefont
  {Steele}}, \bibinfo {author} {\bibfnamefont {B.}~\bibnamefont {Knuffman}},
  \bibinfo {author} {\bibfnamefont {K.~A.}\ \bibnamefont {Twedt}}, \bibinfo
  {author} {\bibfnamefont {A.}~\bibnamefont {Schwarzkopf}}, \ and\ \bibinfo
  {author} {\bibfnamefont {T.~M.}\ \bibnamefont {Wilson}},\ }\href@noop {}
  {\bibfield  {journal} {\bibinfo  {journal} {Appl. Phys. Rev.}\ }\textbf
  {\bibinfo {volume} {3}},\ \bibinfo {pages} {011302} (\bibinfo {year}
  {2016})}\BibitemShut {NoStop}%
\bibitem [{\citenamefont {Franssen}\ \emph {et~al.}(2018)\citenamefont
  {Franssen}, \citenamefont {Kromwijk}, \citenamefont {Vredenbregt},\ and\
  \citenamefont {Luiten}}]{fkv18}%
  \BibitemOpen
  \bibfield  {author} {\bibinfo {author} {\bibfnamefont {J.~G.}\ \bibnamefont
  {Franssen}}, \bibinfo {author} {\bibfnamefont {J.~M.}\ \bibnamefont
  {Kromwijk}}, \bibinfo {author} {\bibfnamefont {E.~J.}\ \bibnamefont
  {Vredenbregt}}, \ and\ \bibinfo {author} {\bibfnamefont {O.~J.}\ \bibnamefont
  {Luiten}},\ }\bibfield  {title} {\enquote {\bibinfo {title} {{Energy spread
  of ultracold electron bunches extracted from a laser cooled gas}},}\
  }\href@noop {} {\bibfield  {journal} {\bibinfo  {journal} {J. Phys. B}\
  }\textbf {\bibinfo {volume} {51}},\ \bibinfo {pages} {035007} (\bibinfo
  {year} {2018})}\BibitemShut {NoStop}%
\bibitem [{\citenamefont {Franssen}\ \emph {et~al.}(2019)\citenamefont
  {Franssen}, \citenamefont {de~Raadt}, \citenamefont {van Ninhuijs},\ and\
  \citenamefont {Luiten}}]{frn19}%
  \BibitemOpen
  \bibfield  {author} {\bibinfo {author} {\bibfnamefont {J.~G.~H.}\
  \bibnamefont {Franssen}}, \bibinfo {author} {\bibfnamefont {T.~C.~H.}\
  \bibnamefont {de~Raadt}}, \bibinfo {author} {\bibfnamefont {M.~A.~W.}\
  \bibnamefont {van Ninhuijs}}, \ and\ \bibinfo {author} {\bibfnamefont
  {O.~J.}\ \bibnamefont {Luiten}},\ }\href@noop {} {\bibfield  {journal}
  {\bibinfo  {journal} {Phys. Rev. Accel. Beams}\ }\textbf {\bibinfo {volume}
  {22}},\ \bibinfo {pages} {023401} (\bibinfo {year} {2019})}\BibitemShut
  {NoStop}%
\bibitem [{\citenamefont {Sadeghi}\ and\ \citenamefont {Grant}(2012)}]{sgr12}%
  \BibitemOpen
  \bibfield  {author} {\bibinfo {author} {\bibfnamefont {H.}~\bibnamefont
  {Sadeghi}}\ and\ \bibinfo {author} {\bibfnamefont {E.~R.}\ \bibnamefont
  {Grant}},\ }\href {\doibase 10.1103/PhysRevA.86.052701} {\bibfield  {journal}
  {\bibinfo  {journal} {Phys. Rev. A}\ }\textbf {\bibinfo {volume} {86}},\
  \bibinfo {pages} {052701} (\bibinfo {year} {2012})}\BibitemShut {NoStop}%
\bibitem [{\citenamefont {Haenel}\ \emph {et~al.}(2017)\citenamefont {Haenel},
  \citenamefont {Schulz-Weiling}, \citenamefont {Sous}, \citenamefont
  {Sadeghi}, \citenamefont {Aghigh}, \citenamefont {Melo}, \citenamefont
  {Keller},\ and\ \citenamefont {Grant}}]{hss17}%
  \BibitemOpen
  \bibfield  {author} {\bibinfo {author} {\bibfnamefont {R.}~\bibnamefont
  {Haenel}}, \bibinfo {author} {\bibfnamefont {M.}~\bibnamefont
  {Schulz-Weiling}}, \bibinfo {author} {\bibfnamefont {J.}~\bibnamefont
  {Sous}}, \bibinfo {author} {\bibfnamefont {H.}~\bibnamefont {Sadeghi}},
  \bibinfo {author} {\bibfnamefont {M.}~\bibnamefont {Aghigh}}, \bibinfo
  {author} {\bibfnamefont {L.}~\bibnamefont {Melo}}, \bibinfo {author}
  {\bibfnamefont {J.~S.}\ \bibnamefont {Keller}}, \ and\ \bibinfo {author}
  {\bibfnamefont {E.~R.}\ \bibnamefont {Grant}},\ }\href@noop {} {\bibfield
  {journal} {\bibinfo  {journal} {Phys. Rev. A}\ }\textbf {\bibinfo {volume}
  {96}},\ \bibinfo {pages} {023613} (\bibinfo {year} {2017})}\BibitemShut
  {NoStop}%
\bibitem [{\citenamefont {Langin}, \citenamefont {Gorman},\ and\ \citenamefont
  {Killian}(2019)}]{lgk19}%
  \BibitemOpen
  \bibfield  {author} {\bibinfo {author} {\bibfnamefont {T.~K.}\ \bibnamefont
  {Langin}}, \bibinfo {author} {\bibfnamefont {G.~M.}\ \bibnamefont {Gorman}},
  \ and\ \bibinfo {author} {\bibfnamefont {T.~C.}\ \bibnamefont {Killian}},\
  }\href@noop {} {\bibfield  {journal} {\bibinfo  {journal} {Science}\ }\textbf
  {\bibinfo {volume} {363}},\ \bibinfo {pages} {61--64} (\bibinfo {year}
  {2019})}\BibitemShut {NoStop}%
\bibitem [{\citenamefont {Mansbach}\ and\ \citenamefont {Keck}(1969)}]{mke69}%
  \BibitemOpen
  \bibfield  {author} {\bibinfo {author} {\bibfnamefont {P.}~\bibnamefont
  {Mansbach}}\ and\ \bibinfo {author} {\bibfnamefont {J.}~\bibnamefont
  {Keck}},\ }\href@noop {} {\bibfield  {journal} {\bibinfo  {journal} {{Phys.
  Rev.}}\ }\textbf {\bibinfo {volume} {181}},\ \bibinfo {pages} {275} (\bibinfo
  {year} {1969})}\BibitemShut {NoStop}%
\bibitem [{\citenamefont {Killian}\ \emph {et~al.}(2001)\citenamefont
  {Killian}, \citenamefont {Lim}, \citenamefont {Kulin}, \citenamefont {Dumke},
  \citenamefont {Bergeson},\ and\ \citenamefont {Rolston}}]{klk01}%
  \BibitemOpen
  \bibfield  {author} {\bibinfo {author} {\bibfnamefont {T.~C.}\ \bibnamefont
  {Killian}}, \bibinfo {author} {\bibfnamefont {M.~J.}\ \bibnamefont {Lim}},
  \bibinfo {author} {\bibfnamefont {S.}~\bibnamefont {Kulin}}, \bibinfo
  {author} {\bibfnamefont {R.}~\bibnamefont {Dumke}}, \bibinfo {author}
  {\bibfnamefont {S.~D.}\ \bibnamefont {Bergeson}}, \ and\ \bibinfo {author}
  {\bibfnamefont {S.~L.}\ \bibnamefont {Rolston}},\ }\href@noop {} {\bibfield
  {journal} {\bibinfo  {journal} {{ Phys. Rev. Lett.}}\ }\textbf {\bibinfo
  {volume} {86}},\ \bibinfo {pages} {3759} (\bibinfo {year}
  {2001})}\BibitemShut {NoStop}%
\bibitem [{\citenamefont {Gupta}\ \emph {et~al.}(2007)\citenamefont {Gupta},
  \citenamefont {Laha}, \citenamefont {Simien}, \citenamefont {Gao},
  \citenamefont {Castro}, \citenamefont {Killian},\ and\ \citenamefont
  {Pohl}}]{gls07}%
  \BibitemOpen
  \bibfield  {author} {\bibinfo {author} {\bibfnamefont {P.}~\bibnamefont
  {Gupta}}, \bibinfo {author} {\bibfnamefont {S.}~\bibnamefont {Laha}},
  \bibinfo {author} {\bibfnamefont {C.~E.}\ \bibnamefont {Simien}}, \bibinfo
  {author} {\bibfnamefont {H.}~\bibnamefont {Gao}}, \bibinfo {author}
  {\bibfnamefont {J.}~\bibnamefont {Castro}}, \bibinfo {author} {\bibfnamefont
  {T.~C.}\ \bibnamefont {Killian}}, \ and\ \bibinfo {author} {\bibfnamefont
  {T.}~\bibnamefont {Pohl}},\ }\href@noop {} {\bibfield  {journal} {\bibinfo
  {journal} {{ Phys. Rev. Lett.}}\ }\textbf {\bibinfo {volume} {99}},\ \bibinfo
  {pages} {75005} (\bibinfo {year} {2007})}\BibitemShut {NoStop}%
\bibitem [{\citenamefont {Perry}\ and\ \citenamefont {Mourou}(1994)}]{pmo94}%
  \BibitemOpen
  \bibfield  {author} {\bibinfo {author} {\bibfnamefont {M.~D.}\ \bibnamefont
  {Perry}}\ and\ \bibinfo {author} {\bibfnamefont {G.}~\bibnamefont {Mourou}},\
  }\href@noop {} {\bibfield  {journal} {\bibinfo  {journal} {{Science}}\
  }\textbf {\bibinfo {volume} {264}},\ \bibinfo {pages} {917} (\bibinfo {year}
  {1994})}\BibitemShut {NoStop}%
\bibitem [{\citenamefont {Raab}\ \emph {et~al.}(1987)\citenamefont {Raab},
  \citenamefont {Prentiss}, \citenamefont {Cable}, \citenamefont {Chu},\ and\
  \citenamefont {Pritchard}}]{rpc87}%
  \BibitemOpen
  \bibfield  {author} {\bibinfo {author} {\bibfnamefont {E.~L.}\ \bibnamefont
  {Raab}}, \bibinfo {author} {\bibfnamefont {M.}~\bibnamefont {Prentiss}},
  \bibinfo {author} {\bibfnamefont {A.}~\bibnamefont {Cable}}, \bibinfo
  {author} {\bibfnamefont {S.}~\bibnamefont {Chu}}, \ and\ \bibinfo {author}
  {\bibfnamefont {D.~E.}\ \bibnamefont {Pritchard}},\ }\href@noop {} {\bibfield
   {journal} {\bibinfo  {journal} {{ Phys. Rev. Lett.}}\ }\textbf {\bibinfo
  {volume} {59}},\ \bibinfo {pages} {2631} (\bibinfo {year}
  {1987})}\BibitemShut {NoStop}%
\bibitem [{\citenamefont {Nagel}\ \emph {et~al.}(2003)\citenamefont {Nagel},
  \citenamefont {Simien}, \citenamefont {Laha}, \citenamefont {Gupta},
  \citenamefont {Ashoka},\ and\ \citenamefont {Killian}}]{nsl03}%
  \BibitemOpen
  \bibfield  {author} {\bibinfo {author} {\bibfnamefont {S.~B.}\ \bibnamefont
  {Nagel}}, \bibinfo {author} {\bibfnamefont {C.~E.}\ \bibnamefont {Simien}},
  \bibinfo {author} {\bibfnamefont {S.}~\bibnamefont {Laha}}, \bibinfo {author}
  {\bibfnamefont {P.}~\bibnamefont {Gupta}}, \bibinfo {author} {\bibfnamefont
  {V.~S.}\ \bibnamefont {Ashoka}}, \ and\ \bibinfo {author} {\bibfnamefont
  {T.~C.}\ \bibnamefont {Killian}},\ }\href@noop {} {\bibfield  {journal}
  {\bibinfo  {journal} {{ Phys. Rev. A}}\ }\textbf {\bibinfo {volume} {67}},\
  \bibinfo {eid} {011401(R)} (\bibinfo {year} {2003})}\BibitemShut {NoStop}%
\bibitem [{\citenamefont {Castro}, \citenamefont {Gao},\ and\ \citenamefont
  {Killian}(2008)}]{cgk08}%
  \BibitemOpen
  \bibfield  {author} {\bibinfo {author} {\bibfnamefont {J.}~\bibnamefont
  {Castro}}, \bibinfo {author} {\bibfnamefont {H.}~\bibnamefont {Gao}}, \ and\
  \bibinfo {author} {\bibfnamefont {T.~C.}\ \bibnamefont {Killian}},\
  }\href@noop {} {\bibfield  {journal} {\bibinfo  {journal} {Plasma Phys.
  Control. Fusion}\ }\textbf {\bibinfo {volume} {50}},\ \bibinfo {pages}
  {124011} (\bibinfo {year} {2008})}\BibitemShut {NoStop}%
\bibitem [{\citenamefont {Langin}\ \emph {et~al.}(2016)\citenamefont {Langin},
  \citenamefont {Strickler}, \citenamefont {Maksimovic}, \citenamefont
  {McQuillen}, \citenamefont {Pohl}, \citenamefont {Vrinceanu},\ and\
  \citenamefont {Killian}}]{lsm16}%
  \BibitemOpen
  \bibfield  {author} {\bibinfo {author} {\bibfnamefont {T.~K.}\ \bibnamefont
  {Langin}}, \bibinfo {author} {\bibfnamefont {T.}~\bibnamefont {Strickler}},
  \bibinfo {author} {\bibfnamefont {N.}~\bibnamefont {Maksimovic}}, \bibinfo
  {author} {\bibfnamefont {P.}~\bibnamefont {McQuillen}}, \bibinfo {author}
  {\bibfnamefont {T.}~\bibnamefont {Pohl}}, \bibinfo {author} {\bibfnamefont
  {D.}~\bibnamefont {Vrinceanu}}, \ and\ \bibinfo {author} {\bibfnamefont
  {T.~C.}\ \bibnamefont {Killian}},\ }\href@noop {} {\bibfield  {journal}
  {\bibinfo  {journal} {Phys. Rev. E}\ }\textbf {\bibinfo {volume} {93}},\
  \bibinfo {pages} {023201} (\bibinfo {year} {2016})}\BibitemShut {NoStop}%
\bibitem [{\citenamefont {McQuillen}\ \emph
  {et~al.}(2015{\natexlab{b}})\citenamefont {McQuillen}, \citenamefont
  {Strickler}, \citenamefont {Langin},\ and\ \citenamefont {Killian}}]{msl15}%
  \BibitemOpen
  \bibfield  {author} {\bibinfo {author} {\bibfnamefont {P.}~\bibnamefont
  {McQuillen}}, \bibinfo {author} {\bibfnamefont {T.}~\bibnamefont
  {Strickler}}, \bibinfo {author} {\bibfnamefont {T.}~\bibnamefont {Langin}}, \
  and\ \bibinfo {author} {\bibfnamefont {T.~C.}\ \bibnamefont {Killian}},\
  }\href@noop {} {\bibfield  {journal} {\bibinfo  {journal} {Phys. Plasmas}\
  }\textbf {\bibinfo {volume} {22}},\ \bibinfo {pages} {033513} (\bibinfo
  {year} {2015}{\natexlab{b}})},\ \Eprint {http://arxiv.org/abs/1501.0694}
  {1501.0694} \BibitemShut {NoStop}%
\bibitem [{\citenamefont {Pohl}, \citenamefont {Vrinceanu},\ and\ \citenamefont
  {Sadeghpour}(2008)}]{pvs08}%
  \BibitemOpen
  \bibfield  {author} {\bibinfo {author} {\bibfnamefont {T.}~\bibnamefont
  {Pohl}}, \bibinfo {author} {\bibfnamefont {D.}~\bibnamefont {Vrinceanu}}, \
  and\ \bibinfo {author} {\bibfnamefont {H.~R.}\ \bibnamefont {Sadeghpour}},\
  }\href@noop {} {\bibfield  {journal} {\bibinfo  {journal} {Phys. Rev. Lett.}\
  }\textbf {\bibinfo {volume} {100}},\ \bibinfo {pages} {223201} (\bibinfo
  {year} {2008})}\BibitemShut {NoStop}%
\bibitem [{\citenamefont {Hamaguchi}, \citenamefont {Farouki},\ and\
  \citenamefont {Dubin}(1997)}]{hfd97}%
  \BibitemOpen
  \bibfield  {author} {\bibinfo {author} {\bibfnamefont {S.}~\bibnamefont
  {Hamaguchi}}, \bibinfo {author} {\bibfnamefont {R.~T.}\ \bibnamefont
  {Farouki}}, \ and\ \bibinfo {author} {\bibfnamefont {D.~H.~E.}\ \bibnamefont
  {Dubin}},\ }\href@noop {} {\bibfield  {journal} {\bibinfo  {journal} {{Phys.
  Rev. E}}\ }\textbf {\bibinfo {volume} {56}},\ \bibinfo {pages} {4671}
  (\bibinfo {year} {1997})}\BibitemShut {NoStop}%
\bibitem [{\citenamefont {Vikhrov}\ \emph {et~al.}(2020)\citenamefont
  {Vikhrov}, \citenamefont {Bronin}, \citenamefont {Klayrfeld}, \citenamefont
  {Zelener},\ and\ \citenamefont {Zelener}}]{vbk20}%
  \BibitemOpen
  \bibfield  {author} {\bibinfo {author} {\bibfnamefont {E.~V.}\ \bibnamefont
  {Vikhrov}}, \bibinfo {author} {\bibfnamefont {S.~Y.}\ \bibnamefont {Bronin}},
  \bibinfo {author} {\bibfnamefont {A.~B.}\ \bibnamefont {Klayrfeld}}, \bibinfo
  {author} {\bibfnamefont {B.~B.}\ \bibnamefont {Zelener}}, \ and\ \bibinfo
  {author} {\bibfnamefont {B.~V.}\ \bibnamefont {Zelener}},\ }\href@noop {}
  {\bibfield  {journal} {\bibinfo  {journal} {Phys. Plasmas}\ }\textbf
  {\bibinfo {volume} {27}},\ \bibinfo {pages} {120702} (\bibinfo {year}
  {2020})}\BibitemShut {NoStop}%
\bibitem [{\citenamefont {Dharodi}\ and\ \citenamefont
  {Murillo}(2020)}]{dvm20}%
  \BibitemOpen
  \bibfield  {author} {\bibinfo {author} {\bibfnamefont {V.~S.}\ \bibnamefont
  {Dharodi}}\ and\ \bibinfo {author} {\bibfnamefont {M.~S.}\ \bibnamefont
  {Murillo}},\ }\href@noop {} {\bibfield  {journal} {\bibinfo  {journal} {Phys.
  Rev. E}\ }\textbf {\bibinfo {volume} {101}},\ \bibinfo {pages} {023207}
  (\bibinfo {year} {2020})}\BibitemShut {NoStop}%
\end{thebibliography}%

\end{document}